\documentclass{optica-article}

\journal{opticajournal} 

\articletype{Research Article}

\usepackage{lineno}
\usepackage{cleveref}
\usepackage{subfigure}
\usepackage{siunitx}

\begin{document}

\title{Beam quality $M^2(\psi)$ factor, spot rotation angle, and angular speed in general laser beams}

\author{Zhen-Xiang Hao,\authormark{1,2}$^\dagger$ Ruo-Xi Wu,\authormark{1,2} Hong-Bo Jin,\authormark{3,4}$^\ast$ Ya-Zheng Tao,\authormark{1,2} and Yue-Liang Wu\authormark{1,2,4,5}}

\address{\authormark{1}School of Fundamental Physics and Mathematical Sciences, Hangzhou Institute for Advanced Study, UCAS, Hangzhou 310024, China\\
\authormark{2}University of Chinese Academy of Sciences, Beijing 100049, China\\
\authormark{3}National Astronomical Observatories, Chinese Academy of Sciences, Beijing 100101, China\\
\authormark{4}The International Center for Theoretical Physics Asia-Pacific, University of Chinese Academy of Sciences, Beijing 100190, China\\
\authormark{5}Institute of Theoretical Physics, Chinese Academy of Sciences, Beijing 100190, China}

\email{\authormark{$\dagger$}haozhenxiang@ucas.ac.cn}
\email{\authormark{$\ast$}hbjin@bao.ac.cn}  

\begin{abstract*} 
A unified definition for the rotation angle and rotation angular speed of general beams, including those with orbital angular momentum (OAM), has been lacking until now. The rotation of a general beam is characterized by observing the rotational behavior of the directions of the extreme spot sizes during propagation. We introduce the beam quality $M^2(\psi)$ factor to characterize the unique beam quality of a general beam across all directions, not limited to the $x$- or $y$-axes. Besides that, we present the beam center $s_{\psi}(\psi,z)$, spot size $w_{\psi}(\psi,z)$, waist position, waist radius, and divergence angle along the direction that forms an angle $\psi$ with the $x$-axis in the plane perpendicular to the $z$-axis for the general beam. Furthermore, this paper presents rapid calculation formulas for these parameters, utilizing the mode expansion method (MEM). Subsequently, we prove that only two extreme spot sizes exist in a given detection plane and the angle between the maximum and minimum spot angles is consistently $90^{\circ}$ during the propagation. We also prove the spot rotation angles converge as $z$ approaches either positive or negative infinity. We first show the extreme spot sizes, spot rotation angle, and angular speed for the vortex beam. Our formulas efficiently differentiate between vortex OAM beams and asymmetry OAM beams.

\end{abstract*}

\section{Introduction}
Since Allen et al. \cite{1992PhRvA..45.8185A} demonstrated in 1992 that Laguerre-Gaussian (LG) beams carry OAM, this discovery has attracted extensive interest in the field. It has ushered modern optics into a new epoch of advancement. Research has shown that laser beams with OAM have significant potential in various applications. For example, they are used in optical trapping and manipulation of microscopic particles \cite{2003Natur.424..810G}, manipulation of cold atoms \cite{1997PhRvL..78.4713K}, phase contrast microscopy \cite{2006OExpr..14.3792B}, and stimulated emission depletion microscopy \cite{2006Natur.440..935W}. Additionally, they are crucial in data transmission through optical fibers \cite{Pryamikov2020}, second-harmonic generation \cite{2020OExpr..28.2536D}, and progress in both classical \cite{2012NaPho...6..488W} and quantum  \cite{2001Natur.412..313M} informatics. Their applications also span secure communication through cryptography \cite{2019OExpr..2731456H}, probing turbulent media \cite{2020OExpr..28..924W}, and even extending into the realm of astronomy \cite{2005OptL...30.3308F}. 

Optical beams can carry OAM in two forms: vortex OAM (VOAM) and asymmetry OAM (AOAM) \cite{2003JOSAA..20.1635B}. VOAM is inherent to optical vortex beams, such as Laguerre-Gaussian (LG) beams, which exhibit wavefront singularities\cite{2001PrOpt..42..219S}. 
AOAM, on the other hand, is associated with the discrepancy between the principal axes of phase and amplitude transverse profiles \cite{1999PrOpt..39..291A}, as seen in general astigmatic Gaussian beams (GAGB).
Under certain conditions, these two forms of OAM can be interconverted \cite{2002JETPL..75..127B}.

During propagation, the spot of a VOAM beam remains stationary, while the spot of an AOAM beam rotates. VOAM beams can theoretically have any integer OAM value, but practical applications often limit it to lower values due to system constraints. In contrast, AOAM beams allow continuous tuning of OAM across a broader range by simply adjusting the cylindrical lens orientation. allows for  anon-integer and greater torque to be applied, enabling various optical manipulations and applications \cite{1997OptCo.144..210C}. Thus, rotating beams like the GAGB have significant application potential, making it essential to evaluate their beam quality and rotational properties.

Laser beam characterization is widely used in laser theoretical analysis, design, medical treatment, laser welding, and many other fields\cite{s16122014}. The quality of a laser beam is a crucial metric for assessing its performance and is highly significant in both theoretical and applied laser research \cite{1998ApOpt..37.4840J,2004ApOpt..43.5037L,2007ApPhL..90m1105D}. Various definitions of beam quality are employed across different scenarios, such as power spectral density (PSD) \cite{1995ApOpt..34..201E}, Strehl ratio \cite{Mahajan:82}, Power in the bucket (PIB) \cite{KANG20061}, beam quality $M^2$ factor \cite{s16122014} and so on. The most widely used definition of beam quality is the beam quality $M^2$ factor proposed by Siegman \cite{1990SPIE.1224....2S}. The International Organization for Standardization (ISO) recommends the beam quality $M^2$ factor as an important criterion for evaluating beam quality due to its universality, ease of measurement, and invariant propagation through the ABCD system \cite{ISO11146-1, ISO11146-2}.

The traditional $M_x^2$ and $M_y^2$ are limited to evaluating the beam quality along the $x$-axis and $y$-axis, respectively. When the coordinate system is rotated, $M_x^2$ and $M_y^2$ values may fluctuate, showing their dependency on the chosen coordinate system. To address this limitation, innovative definitions such as the beam quality $M^2$ matrix \cite{2009LaPhy..19..455L} and $M^2$-curve \cite{s16122014} have been introduced for characterizing asymmetric beams. These new metrics can resolve the non-uniqueness issue associated with the traditional beam quality $M^2$ factor in certain special cases. However, they do not reveal information about the inherent rotational motion within AOAM beams.

In recent years, multimode fiber has attracted significant attention in the telecommunications field. Advanced fast mode decomposition methods have been developed to enable faster and more detailed analysis of these fibers \cite{2020NatCo..11.5507M,2009OExpr..17.9347K}. These methods have led to the development of techniques for real-time assessment of the beam quality $M^2$ factor using the MEM \cite{2011OExpr..19.6741S}.
The beam quality factor, denoted as $M^2$, is determined by both the spot size and the divergence angle. In our previous work \cite{2023IJMPD..3250134H}, we refined the formula for the mean square deviation (MSD) spot size of the MEMS beam, facilitating rapid calculation of the spot size for any propagation distance. Building on this foundation, this paper introduces rapid calculation formulas for the direction, divergence angle, waist position, waist radius, and $M^2$ factors in both the $x$- and $y$- directions of the MEMS beam.
Additionally, we introduce several new definitions: the beam quality factor $M^2(\psi)$, the spot rotation angle $\psi_{e}$, and the spot rotation angular speed $\psi_{e}'$. The $M^2(\psi)$ factor, a novel definition of beam quality, characterizes the unique beam quality across all directions, not confined to the $x$- or $y$-axes. The spot rotation angle $\psi_{e}$ and the spot rotation angular speed $\psi_{e}'$ clarify the rotational motion characteristics of the beam.
We also present the beam center $s_{\psi}(\psi,z)$, spot size $w_{\psi}(\psi,z)$, waist position, waist radius, and divergence angle along the direction forming an angle $\psi$ with the $x$-axis in the plane perpendicular to the $z$-axis for the general beam. This paper provides rapid calculation formulas for these parameters, utilizing the MEMS technology.
We demonstrate that there are only two extreme spot sizes in a given detection plane, with the angle between the maximum and minimum spot sizes consistently being $90^{\circ}$ during propagation. We also show that the spot rotation angles converge as $z$ approaches positive or negative infinity.
We first illustrate the extreme spot sizes, spot rotation angle, and angular speed for the vortex beam, such as the elliptical vHG beam. Our formulas effectively differentiate between VOAM beams and AOAM beams.

In Section \cref{dirver_mem_msd}, we review the traditional beam quality $M^2$ factor and derive rapid calculation formulas for the direction, divergence angle, waist position, waist size, and beam quality $M^2$ factors in both the $x$- and $y$-directions for the general MEMS beam. We then introduce new definitions for the beam quality $M^2(\psi)$ factor, the spot rotation angle $\psi_{e}$, and the spot rotation angular speed $\psi_{e}'$ in Section \cref{best_spot}, where we also present rapid calculation formulas for these parameters using MEMS technology. Subsequently, in Section \cref{appli}, we evaluate the performance of these formulas across various scenarios, including the rotated simple astigmatic Gaussian beam, the oblique high-order HG beam, the GAGB, and the vortex Hermite-Gaussian beam. Finally, we conclude the research with a summary of our findings.

\section{Direction, Divergence Angle, Waist Position, Waist Radius, And The Beam Quality $M^2$ Factor For The General beams}\label{dirver_mem_msd}
 
 The incoherent superposition of Hermite-Gaussian (HG) modes can be used to describe the output beam of a multimode laser operating within a stable cavity\cite{s16122014}. In MEM, the complex amplitude of the general beam is approximated by the superposition of a finite number of HG modes\cite{2023IJMPD..3250134H}. This is expressed as: 
\begin{equation}
u(x,y,z)\approx\sum_{m=0}^N\sum_{n=0}^{N-m} a_{mn}u_{mn}(x,y,z),
\end{equation}
 where the maximum mode order number $N$ represents the truncated order number, $u_{mn}(x,y,z)$ denotes the complex amplitude of the HG mode for the order $(m,n)$, and $a_{mn}$ signifies the corresponding complex mode amplitude. These basic HG modes share the same waist, direction, and beam center. We assume that all HG modes propagate along the $z$-axis, with $x$ and $y$ representing the Cartesian coordinates in the cross-sectional plane of these modes. For simplicity, here we use the symmetric HG modes as decomposition bases. The formula for $u_{mn}(x,y,z)$ in its local coordinate system is given by
 \begin{equation}
 \begin{split}
 	u_{mn}(x,y,z)=&\frac{c_{mn}}{w(z)}H_m\left(\frac{\sqrt{2}x}{w(z)}\right)H_n\left(\frac{\sqrt{2}y}{w(z)}\right)\exp{\left(-\frac{x^2+y^2}{w^2(z)}\right)}\\
	&\cdot\exp{\left(-ikz-ik\frac{x^2+y^2}{2R(z)}+i(m+n+1)\zeta(z)\right)},
	\label{hmn_complex_amp} 
\end{split}	
 \end{equation}
where $c_{mn}=(\pi m!n!2^{m+n-1})^{-\frac{1}{2}}$ is the normalization constant, $R(z)=z(1+({z_r}/{z})^2)$ represents the radius of curvature for the fundamental Gaussian mode $u_{00}(x,y,z)$, and $\zeta(z)=\arctan({z}/{z_r})$ denotes the corresponding Gouy phase for $u_{00}(x,y,z)$. $H_m$ and $H_n$ denotes the Hermite-polynomial of order $m$ in the $x$-direction and $n$ in the $y$-direction, respectively. The parameter $w(z)=w_0\sqrt{1+({z}/{z_r})^2}$ signifies the spot size of $u_{00}(x,y,z)$, where $w_0$ is the waist, $z_r={\pi{w_0}^2}/{\lambda}$ is the Rayleigh range, $\lambda$ is the wavelength, and $k={2\pi}/{\lambda}$ is the wave number.  In this paper, the term "MEM beam" refers to the general beam that has been decomposed by the MEM. For the sake of simplicity, we conduct all the calculations in this article within the local coordinate system of $u_{mn}(x,y,z)$.

The traditional MSD spot size is defined as \cite{1992ApOpt..31.6389P}
\begin{equation}
w_j(z)=2\sqrt{\frac{\iint_{-\infty}^{+\infty}(j-s_j(z))^2I(x,y,z)dxdy}{P_{in}}}=2\sqrt{\frac{\iint_{-\infty}^{+\infty}j^2I(x,y,z)dxdy}{P_{in}}-s_j(z)^2}.\label{spot_integ}
\end{equation}

where $I(x,y,z)$ represents the intensity of the beam, $P_{in}$ denotes the total power of the beam, and $j$ signifies either $x$ or $y$, coresponding to the $x$- or $y$-direction, repectively. The beam center $s_j(z)$ is defined as the mean value of the transversal position of the beam in one dimension \cite{1992ApOpt..31.6389P}:
\begin{equation}
s_j(z)=\frac{\iint_{-\infty}^{+\infty}jI(x,y,z)dxdy}{P_{in}}.\label{center_integ}
\end{equation}

If one utilizes \cref{spot_integ} and \cref{center_integ} to calculate the spot size and beam center of a general beam, different integrals must be solved for various propagation distances. However, by employing the MEM to decompose general beams, we can analytically solve the integrals within the definitions of the MSD spot size and beam center. Consequently, people are no longer required to compute these separate integrals for various propagation distances. This approach significantly reduces the computational cost associated with estimating the spot size and beam center, particularly for complex applications\cite{2023IJMPD..3250134H}.
 
Building on this foundation, we further aim to provide formulas that enable rapid calculation of key parameters for a general beam. These parameters include the beam propagation direction, waist radius, waist position, divergence angle, and beam quality $M^2$ factor. For the sake of brevity, we adopt the same notation as utilized in Ref.\cite{2023IJMPD..3250134H}
\begin{equation}
u_{mn}(x,y,z)=u_{mn}=|u_{mn}|\exp{\left(i\phi_{mn}\right)},
\end{equation}
\begin{equation}
a_{mn}=|a_{mn}|\exp{(i\beta_{mn})}=\iint_{-\infty}^{+\infty}u(x,y,z)u^*_{mn}(x,y,z)dxdy.
\end{equation}

The formulas for the spot size and beam center, as given by eq.(11) and eq.(10) in Ref.\cite{2023IJMPD..3250134H}, are functions of $z$. However, the independent variable $z$ is implicitly contained within in $w(z)$ and $\zeta(z)$. Here we explicitly derive the relationships between the beam center $(s_x(z),s_y(z))$ and $z$:
\begin{equation}
\begin{split}
s_x(z)=&\frac{w_0}{P_{MEM}}\sum_{m=0}^{N}\sum_{n=0}^{N-m}|a_{mn}||a_{(m+1)n}|\sqrt{m+1}\\
&\cdot\left(\cos{(\beta_{mn}-\beta_{(m+1)n})}+\sin{(\beta_{mn}-\beta_{(m+1)n})}\frac{z}{z_r}\right),
\end{split}
\end{equation}
\begin{equation}
\begin{split}
s_y(z)=&\frac{w_0}{P_{MEM}}\sum_{m=0}^{N}\sum_{n=0}^{N-m}|a_{mn}||a_{m(n+1)}|\sqrt{n+1}\\
&\cdot\left(\cos{(\beta_{mn}-\beta_{m(n+1)})}+\sin{(\beta_{mn}-\beta_{m(n+1)})}\frac{z}{z_r}\right).
\end{split}
\end{equation}

It is evident that the beam center $(s_x(z),s_y(z))$ is a linear function of $z$. The slope of this linear function can be used to define the propagation direction $\vec{dir}$ of the MEM beam:
\begin{equation}
\begin{split}
\vec{dir}=&\left(\frac{w_0}{P_{MEM}z_r}\sum_{m=0}^{N}\sum_{n=0}^{N-m}t_{mn}\sin{\left(\beta_{mn}-\beta_{(m+1)n}\right)},\right.\\
&\left.\frac{w_0}{P_{MEM}z_r}\sum_{m=0}^{N}\sum_{n=0}^{N-m}q_{mn}\sin{\left(\beta_{mn}-\beta_{m(n+1)}\right)},1\right)\label{propa_dir},
\end{split}
\end{equation}
where $t_{mn}=|a_{mn}||a_{(m+1)n}|\sqrt{m+1}$ and $q_{mn}=|a_{mn}||a_{m(n+1)}|\sqrt{n+1}$. A similar one-dimensional expression can be found in Ref.\cite{1992ApOpt..31.6389P}. They utilize the Fourier transform to convert the spatial complex amplitude of the general beam into its spatial frequency complex amplitude. The authors subsequently define the expectation of this new amplitude in the spatial frequency domain as the beam's direction \cite{1992ApOpt..31.6389P}. Our derivation process is considerably more straightforward than theirs.

Similarly, we can explicitly express the relationships between the spot size and $z$: 
\begin{equation}
w_x^2(z)=\frac{w_0^2}{P_{MEM}^2z_r^2}\left(a_xz^2+b_xzz_r+c_xz_r^2\right)\label{eq:wx_z},
\end{equation}

\begin{equation}
w_y^2(z)=\frac{w_0^2}{P_{MEM}^2z_r^2}\left(a_yz^2+b_y z z_r+c_y z_r^2\right)\label{eq:wy_z}, 
\end{equation}
where the specific expression for the coefficients $a_x,b_x,c_x,a_y,b_y,c_y$ are detailed in \cref{appen_coe}. \Cref{spot_integ} shows that the MSD spot size is consistently non-negative. By utilizing this property, we can directly prove that $a_x>0$ and $a_y>0$. Consequently, the aforementioned equations indicate that the MSD spot sizes are the hyperbolic functions of $z$. Using the properties of the hyperbolic functions, we can determine the waist positions $z_{x_{min}}$ for the $x$-direction and $z_{y_{min}}$ for the $y$-directions, respectively:
\begin{equation}
\begin{split}
z_{x_{min}}=&\frac{-b_x}{2a_x}z_r,\\
z_{y_{min}}=&\frac{-b_y}{2a_y}z_r.
\end{split}\label{waist_location}
\end{equation}

The corresponding waist radii $w_{x_{min}}$ for the $x$-direction and $w_{y_{min}}$ for the $y$-directions are given by
\begin{equation}
\begin{split}
w_{x_{min}}=\frac{w_0}{P_{MEM}}\sqrt{\frac{-b_x^2}{4a_x}+c_x},\\
w_{y_{min}}=\frac{w_0}{P_{MEM}}\sqrt{\frac{-b_y^2}{4a_y}+c_y}.
\end{split}
\end{equation}

As is well known, the hyperbolic functions $w_x(z)$ and $w_y(z)$ increase approximately linearly with $z$ when $z\gg 0$. Employing the same definition as for the fundamental Gaussian beam, we can derive the divergence angles for the MEM beam:
\begin{equation}
\begin{split}
\Phi_{MEM_x}=&\frac{w_0}{P_{MEM}z_r}\sqrt{a_x},\\
\Phi_{MEM_y}=&\frac{w_0}{P_{MEM}z_r}\sqrt{a_y}.
\end{split}\label{mem_div_ang}
\end{equation}

A similar one-dimensional expression can also be found in Ref.\cite{1992ApOpt..31.6389P}. Our derivation process is also significantly simpler than theirs. Finally, the beam quality $M^2$ factors in the $x$- and $y$-directions are as follows:

\begin{equation}
\begin{split}
M_x^2=&\frac{\pi}{\lambda}w_{x_{min}}\Phi_{MEM_x}=\frac{1}{P_{MEM}^2}\sqrt{a_xc_x-\frac{b_x^2}{4}},\\
M_y^2=&\frac{\pi}{\lambda}w_{y_{min}}\Phi_{MEM_y}=\frac{1}{P_{MEM}^2}\sqrt{a_yc_y-\frac{b_y^2}{4}}.
\end{split}
\end{equation}

\section{Determination of the beam quality $M^2(\psi)$ factor, spot rotation angle and angular speed}\label{best_spot}

\begin{figure}[htbp]
	\centering
	\includegraphics[width=12cm]{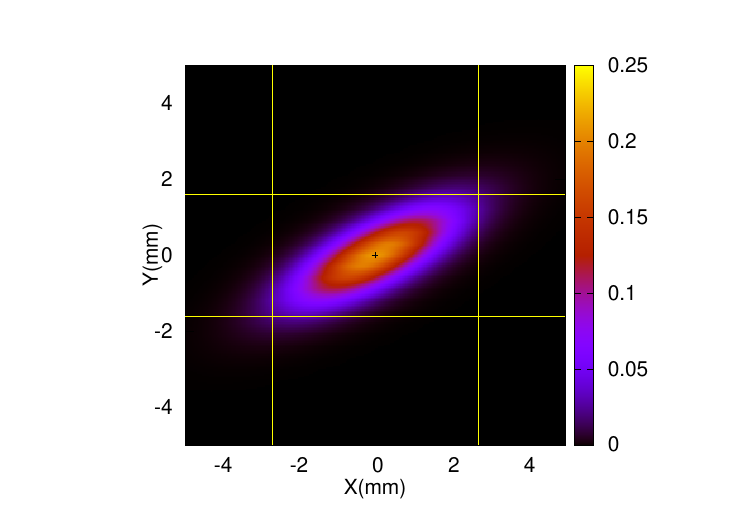}
	\centering
	\caption{The intensity profile of simple astigmatic Gaussian beam. }\label{sagb_example}
\end{figure}

We note that the aforementioned formulas are designed to calculate the spot size along the $x$- or $y$-axis only. For circular beams, such as the fundamental Gaussian beam, the spot size remains consistent across all directions. In this scenario, calculating the spot size along the $x$-direction alone is sufficient. However, for non-circular beams, the spot size along the $x$- or $y$-axis often does not represent the maximum or minimum width within the beam's cross-section. Consider, for instance, a simple astigmatic Gaussian beam; its intensity profile, depicted in \cref{sagb_example}, shows the spot sizes for the $x$- and $y$-directions marked by yellow lines. This figure clearly illustrates that the spot sizes along the $x$- and $ y$-axes are not the extreme spot sizes of the beam. The beam quality $M_x^2$ and $M_y^2$ factors also vary with coordinate axis rotation. Furthermore, beams with rotational intensity profiles, such as the GAGB, which rotate as they propagate, hold significant potential for various applications, making the assessment of their beam quality and rotational characteristics essential.

Analogous to \cref{spot_integ} and \cref{center_integ}, we define the beam center $s_{\psi}(\psi,z)$ and the spot size $w_{\psi}(\psi,z)$  along the direction that forms an angle $\psi$ with the $x$-axis in the plane perpendicular to the $z$-axis:
\begin{equation}
s_{\psi}(\psi,z)=\frac{\iint_{-\infty}^{+\infty} \left(x\cos{\psi}+y\sin{\psi}\right)I(x,y,z)dxdy}{P_{in}} \label{beam_cen_theta_ini},
\end{equation}

\begin{equation}
w_{\psi}(\psi,z)=2\sqrt{\frac{\iint_{-\infty}^{+\infty}  \left(x\cos{\psi}+y\sin{\psi}\right)^2I(x,y,z)dxdy}{P_{in}}-s_{\psi}(\psi,z)^2}\label{spot_theta_ini}.
\end{equation}

By utilizing MEM, we can evaluate the integrals in \cref{beam_cen_theta_ini} and \cref{spot_theta_ini}, and the results are as follows:
\begin{equation}
\begin{split}
s_{\psi}(\psi,z)\approx&\frac{\iint_{-\infty}^{+\infty}  \left(x\cos{\psi}+y\sin{\psi} \right)I(x,y,z)dxdy}{P_{MEM}}\\
=&s_x(z)\cos{\psi}+s_y(z)\sin{\psi}\\
=&\frac{w(z)}{P_{MEM}} \sum_{m=0}^{N}\sum_{n=0}^{N-m}\left(|a_{mn}||a_{(m+1)n}|\cos{(\beta_{mn}-\beta_{(m+1)n}-\zeta(z))}\sqrt{m+1}\cos{\psi}\right.\\
&\left.+|a_{mn}||a_{m(n+1)}|\cos{(\beta_{mn}-\beta_{m(n+1)}-\zeta(z))}\sqrt{n+1}\sin{\psi}\right)\label{bahg_theta},
\end{split}
\end{equation}

\begin{equation}
\begin{split}
&w_{\psi}^2(\psi,z)=\frac{w^2(z)}{P_{MEM}}\left(\sum_{m=0}^{N}\sum_{n=0}^{N-m}|a_{mn}|^2\left(2m\cos^2{\psi}+2n\sin^2{\psi}+1\right)\right.\\
&+2 \sum_{m=0}^{N}\sum_{n=0}^{N-m}|a_{mn}||a_{(m+2)n}|\cos{(\beta_{mn}-\beta_{(m+2)n}-2\zeta(z))}\sqrt{(m+2)(m+1)}\cos^2{\psi}\\
&+2 \sum_{m=0}^{N}\sum_{n=0}^{N-m}|a_{mn}||a_{m(n+2)}|\cos{(\beta_{mn}-\beta_{m(n+2)}-2\zeta(z))}\sqrt{(n+2)(n+1)}\sin^2{\psi}\\
&+4\sum_{m=0}^{N}\sum_{n=0}^{N-m}|a_{mn}||a_{(m+1)(n+1)}|\cos{(\beta_{mn}-\beta_{(m+1)(n+1)}-2\zeta(z))}\\
&\cdot\sqrt{(m+1)(n+1)}\cos{\psi}\sin{\psi}\\
&\left.+4\sum_{m=0}^{N}\sum_{n=0}^{N-m}|a_{m(n+1)}||a_{(m+1)n}|\cos{(\beta_{m(n+1)}-\beta_{(m+1)n})}\sqrt{(m+1)(n+1)}\cos{\psi}\sin{\psi}\right)\\
&-4s_{\psi}(\psi,z)^2\label{wxhg_theta}.
\end{split}
\end{equation}

\Cref{bahg_theta} indicates that the position of the beam center in the direction that forms an angle $\psi$ with the $x$-axis in the plane perpendicular to the $z$-axis is the projection of the beam center $(s_x(z),s_y(z))$. The extremum values of the MSD spot size on the cross-section of the HG modes occur when the derivative of \cref{wxhg_theta} equals zero:

\begin{scriptsize}
\begin{equation}
\begin{split}
\frac{\partial w_{\psi}(\psi,z)^2}{\partial\psi}=&\frac{4w(z)^2}{P_{MEM}^2}\left(\bigl(P_{MEM}\mathcal W_1-2\mathcal X_1\mathcal X_2\bigr)\bigl(\cos^2\psi-\sin^2\psi\bigr)+\bigl(P_{MEM}\mathcal W_2+2\mathcal X_1^2-2\mathcal X_2^2\bigr) \cos\psi\sin\psi \right)=0.
\end{split}\label{wxhg_theta_derive}
\end{equation}
\end{scriptsize}

The expressions of $\mathcal W_1$, $\mathcal W_2$, $\mathcal X_1$ and $\mathcal X_2$ are given in \cref{appen_coe}. In order to solve the above equation, we define
\begin{equation}
\sin{\phi_0}=\frac{2\bigl(P_{MEM}\mathcal W_1-2\mathcal X_1\mathcal X_2\bigr)}{\sqrt{4\bigl(P_{MEM}\mathcal W_1-2\mathcal X_1\mathcal X_2\bigr)^2+\bigl(P_{MEM}\mathcal W_2+2\mathcal X_1^2-2\mathcal X_2^2\bigr)^2}}.
\end{equation}

Subsequently, \cref{wxhg_theta_derive} can be rewritten as
\begin{small}
\begin{equation}
\sqrt{4\bigl(P_{MEM}\mathcal W_1-2\mathcal X_1\mathcal X_2\bigr)^2+\bigl(P_{MEM}\mathcal W_2+2\mathcal X_1^2-2\mathcal X_2^2\bigr)^2}\bigl(\sin{\phi_0}\cos{2\psi}+\cos{\phi_0}\sin{2\psi}\bigr)=0.
\end{equation}
\end{small}

This equation simplifies under the conditions $ P_{MEM}\mathcal W_1-2\mathcal X_1\mathcal X_2\neq 0$ and $P_{MEM}\mathcal W_2+2\mathcal X_1^2-2\mathcal X_2^2\neq0$:
\begin{equation}
\sin{(2\psi+\phi_0)}=0\label{wxhg_theta_derive_simp}.
\end{equation}

This equation indicates that there are exactly 2 extreme values for the MSD spot size on the cross-section of the HG modes: one corresponds to the rotation angle where the spot size is maximized (maximum spot rotation angle) and the other corresponds to the rotation angle where the spot size is minimized (minimum spot rotation angle). The rotation angles corresponding to these extreme values of the spot size are:

\begin{scriptsize}
\begin{equation}
\psi_{e1}=-\frac{1}{2}\phi_0=\left\{
\begin{aligned}
&\pi-\frac{1}{2}\arccos{\frac{P_{MEM}\mathcal W_2+2\mathcal X_1^2-2\mathcal X_2^2}{\sqrt{4\bigl(P_{MEM}\mathcal W_1-2\mathcal X_1\mathcal X_2\bigr)^2+\bigl(P_{MEM}\mathcal W_2+2\mathcal X_1^2-2\mathcal X_2^2\bigr)^2}}}\ ,\ P_{MEM}\mathcal W_1-2\mathcal X_1\mathcal X_2\geq 0,\\
&\frac{1}{2}\arccos{\frac{P_{MEM}\mathcal W_2+2\mathcal X_1^2-2\mathcal X_2^2}{\sqrt{4\bigl(P_{MEM}\mathcal W_1-2\mathcal X_1\mathcal X_2\bigr)^2+\bigl(P_{MEM}\mathcal W_2+2\mathcal X_1^2-2\mathcal X_2^2\bigr)^2}}}\ ,\ P_{MEM}\mathcal W_1-2\mathcal X_1\mathcal X_2<0,
\end{aligned}\label{theta_ex1}
\right.
\end{equation}
\end{scriptsize}
\begin{scriptsize}
\begin{equation}
\psi_{e2}=\frac{\pi}{2}-\frac{1}{2}\phi_0=\left\{
\begin{aligned}
&\frac{3\pi}{2}-\frac{1}{2}\arccos{\frac{P_{MEM}\mathcal W_2+2\mathcal X_1^2-2\mathcal X_2^2}{\sqrt{4\bigl(P_{MEM}\mathcal W_1-2\mathcal X_1\mathcal X_2\bigr)^2+\bigl(P_{MEM}\mathcal W_2+2\mathcal X_1^2-2\mathcal X_2^2\bigr)^2}}}\ ,\ P_{MEM}\mathcal W_1-2\mathcal X_1\mathcal X_2\geq 0,\\
&\frac{\pi}{2}+\frac{1}{2}\arccos{\frac{P_{MEM}\mathcal W_2+2\mathcal X_1^2-2\mathcal X_2^2}{\sqrt{4\bigl(P_{MEM}\mathcal W_1-2\mathcal X_1\mathcal X_2\bigr)^2+\bigl(P_{MEM}\mathcal W_2+2\mathcal X_1^2-2\mathcal X_2^2\bigr)^2}}}\ ,\ P_{MEM}\mathcal W_1-2\mathcal X_1\mathcal X_2< 0.
\end{aligned}\label{theta_ex2}
\right.
\end{equation}
\end{scriptsize}

To ensure that $\psi_{e1}\in(0,\pi)$ and $\psi_{e2}\in(\frac{\pi}{2},\frac{3\pi}{2})$, we add $\pi$ in the case where $P_{MEM}\mathcal W_1-2\mathcal X_1\mathcal X_2\geq 0$. It' is easy to observe that the angle between the maximum spot rotation angle and the minimum spot rotation angle remains constant at $90^{\circ}$ throughout propagation. This implies that their spot rotation angular speeds are always equal. If $ P_{MEM}\mathcal W_1-2\mathcal X_1\mathcal X_2= 0$ and $P_{MEM}\mathcal W_2+2\mathcal X_1^2-2\mathcal X_2^2=0$, then \cref{wxhg_theta_derive} will always be zero. In this situation, the spot size remains consistent across all directions. If $ P_{MEM}\mathcal W_1-2\mathcal X_1\mathcal X_2= 0$ and $P_{MEM}\mathcal W_2+2\mathcal X_1^2-2\mathcal X_2^2\neq0$, then the extreme angles $\psi_{e1}$ and $\psi_{e2}$ are $0,\ \frac{\pi}{2},\  \pi\ $ or $\ \frac{3\pi}{2}$. We refer to $\psi_{e1}$ and $\psi_{e2}$ as the spot rotation angles.

The maximum and minimum spot rotation angles are dependent on $z$. Upon careful observation of \cref{w1}, \cref{w2}, \cref{x1} and \cref{x2}, we note that the spot rotation angles $\psi_{e}$ tend to stabilize as $z$ approaches either positive or negative infinity. Furthermore, by taking the derivative of \cref{theta_ex1} with respect to $z$, we can determine the spot rotation angular speed:

\begin{equation}
\begin{split}
\psi_{e}'=\psi_{e1}'=\psi_{e2}'=\frac{d\psi_{e1}}{dz}=\pm&\left(\frac{(P_{MEM}\mathcal W_2'+4\mathcal X_1\mathcal X_1'-4\mathcal X_2\mathcal X_2')(P_{MEM}\mathcal W_1-2\mathcal X_1\mathcal X_2)}{4\bigl(P_{MEM}\mathcal W_1-2\mathcal X_1\mathcal X_2\bigr)^2+\bigl(P_{MEM}\mathcal W_2+2\mathcal X_1^2-2\mathcal X_2^2\bigr)^2}\right.\\
&\left.-\frac{(P_{MEM}\mathcal W_2+2\mathcal X_1^2-2\mathcal X_2^2)(P_{MEM}\mathcal W_1'-2\mathcal X_1'\mathcal X_2-2\mathcal X_1\mathcal X_2')}{4\bigl(P_{MEM}\mathcal W_1-2\mathcal X_1\mathcal X_2\bigr)^2+\bigl(P_{MEM}\mathcal W_2+2\mathcal X_1^2-2\mathcal X_2^2\bigr)^2}\right),
\end{split}\label{rota_speed}
\end{equation}

where $\mathcal W_1'$,$\mathcal W_2'$,$\mathcal X_1'$, and $\mathcal X_2'$ are detailed in \cref{appen_coe}. The sign of \cref{rota_speed} is positive when $P_{MEM}\mathcal W_1-2\mathcal X_1\mathcal X_2<0$ and the sign is negative when $ P_{MEM}\mathcal W_1-2\mathcal X_1\mathcal X_2\geq 0$. The maximum spot angle, the minimum spot angle, and their rotation angular speed can be used to describe the rotational characteristics of the general beam. We refer to $\psi_{e}'$ as the spot rotation angular speed.

Similarly, by following the approach outlined in the previous section, we will reformulate the expressions for $s_{\psi}(\psi,z)$ and $w_{\psi}(\psi,z)$ to clearly delineate their dependence on $z$ 
\begin{equation}
	s_{\psi}(\psi,z) =\frac{w_0}{P_{MEM}}\bigl( (\mathcal X_{s}(\psi)+\mathcal Y_{s}(\psi)) \frac{z}{z_r}
     +(\mathcal X_{c}(\psi)+\mathcal Y_{c}(\psi)) \bigr), \label{hgX(z)}
\end{equation}

\begin{equation}
w_{\psi}(\psi,z)^2=\frac{w_0^2}{P_{MEM}^2z_r^2}(\mathcal Z_1(\psi)\cdot z^2+\mathcal Z_2(\psi)\cdot z_r z+\mathcal Z_3(\psi)\cdot z_r^2),
\end{equation}
where $\mathcal X_{s}$, $\mathcal Y_{s}$, $\mathcal X_{c}$, $\mathcal Y_{c}$, $\mathcal Z_1$, $\mathcal Z_2$, and$\mathcal Z_3$ are detailed in \cref{appen_coe}. The spot size $w_{\psi}(\psi,z)$ is still a hyperbolic function of $z$.
By utilizing the properties of hyperbolas, we derive the formulas for the waist position $z_{min}(\psi)$ and waist size $w_{min}(\psi)$ along the direction that forms an angle $\psi$ with the $x$-axis in the plane perpendicular to the $z$-axis for the general beams:
\begin{equation}
w_{min}(\psi)=\frac{w_0}{P_{MEM}}\sqrt{\mathcal Z_3(\psi)-\frac{\mathcal Z_2(\psi)^2}{4\mathcal Z_1(\psi)}},
\end{equation}

\begin{equation}
z_{min}(\psi)=\frac{-\mathcal Z_2(\psi)}{2\mathcal Z_1(\psi)}z_r\label{waist_location_theta}.
\end{equation}

The divergence angle $\Phi_{MEM}(\psi)$ along the direction that forms an angle $\psi$ with the $x$-axis in the plane perpendicular to the $z$-axis is given by
\begin{equation}
\Phi_{MEM}(\psi)=\frac{w_0}{P_{MEM}z_r}\sqrt{\mathcal Z_1(\psi)}.
\end{equation}

The beam quality $M^2(\psi)$ factor along the direction that forms an angle $\psi$ with the $x$-axis in the plane perpendicular to the $z$-axis is given by

\begin{equation}
M^2(\psi)=\frac{\pi}{\lambda}w_{min}(\psi)\Phi_{MEM}(\psi)=\frac{1}{P_{MEM}^2}\sqrt{\mathcal Z_3(\psi)\mathcal Z_1(\psi)-\frac{\mathcal Z_2(\psi)^2}{4}}\label{m2curve}.
\end{equation}

The projections of the beam quality $M^2(\psi)$ factor on the $x$- and $y$-axes are given by
\begin{equation}
\begin{split}
M^2(\psi)_x=M^2(\psi)\cos{\psi},\\
M^2(\psi)_y=M^2(\psi)\sin{\psi}.
\end{split}
\end{equation}

\section{Application}\label{appli}

In this section, we will validate the new definitions of the beam quality $M^2(\psi)$ factor, the spot rotation angle $\psi_{e}$, and the corresponding spot rotation angular speed $\psi_{e}'$ through tests involving the rotated simple astigmatic Gaussian beam, the oblique high-order Hermite-Gaussian beam, the general astigmatic gaussian beam, and the elliptical vortex Hermite-Gaussian beam, which are all commonly observed in laboratory experiments. These tests are utilized to calibrate our numerical code. The numerical calculations for the analytical expressions presented in this paper are performed using C++ code with double precision.

\subsection{Rotated Simple Astigmatic Gaussian Beam}

When a Gaussian beam strikes a curved surface at an oblique angle, it transforms into a simple astigmatic Gaussian beam, which has an elliptical spot\cite{Kochkina2013,1969ApOpt...8..975M,alda2003}. We assume that the simple astigmatic Gaussian beam, which propagates in the $+z$ direction, does not have its major axis of the spot ellipse aligned with the $x$-axis. We denote the angle $\theta_s$ as the rotation angle formed between the major axis of the spot ellipse and the $x$-axis. Furthermore, we assume that the beam is offset, with the center of the simple astigmatic Gaussian beam positioned at $(x_{s_0},y_{s_0})$. In this scenario, the original MSD spot sizes $w_x(z)$ and $w_y(z)$, along with their corresponding beam quality $M_x^2$ and $M_y^2$ factors are inconsistent with the theoretical predictions. Here we will demonstrate that our derived extreme spot sizes $w_{\psi}(\psi_e,z)$ and the corresponding beam quality $M^2(\psi_e)$ factors are consistent with the theoretical results.

The complex amplitude of the simple astigmatic Gaussian beam is given by \cite{kochkina2013stigmatic}
\begin{equation}
\begin{split}
u_{s}=&\sqrt{\frac{P}{\pi}}\frac{1}{\sqrt{w_xw_y}}\exp{(-\frac{x_s^2(\theta_s)}{w_x^2}-\frac{y_s^2(\theta_s)}{w_y^2})}\\
&\times\exp{ \left(i\left(-k\left(\frac{x_s^2(\theta_s)}{2R_x}+\frac{y_s^2(\theta_s)}{2R_y}\right)+\frac{1}{2}\left(\zeta_x+\zeta_y\right)-kz\right)\right)},
\end{split}
\end{equation}
\begin{equation}
x_s(\theta_s)=(x-x_{s_0})\cos{\theta_s}+(y-y_{s_0})\sin{\theta_s}\ \ , \ y_s(\theta_s)=-(x-x_{s_0})\sin{\theta_s}+(y-y_{s_0})\cos{\theta_s},
\end{equation}
where $w_{x,y}=w_{0_{x,y}}\sqrt{1+({z}/{z_{r_{x,y}}})^2}$ is the spot size for $x$- or $y$-direction of the simple astigmatic Gaussian beam, $w_{0_{x,y}}$ is the corresponding beam waist, and $z_{r_{x,y}}={\pi w_{0_{x,y}}^2}/{\lambda}$ is the corresponding Rayleigh range. The radius of curvature is $R_{x,y}=z(1+({z_{r_{x,y}}}/{z})^2)$ and the Gouy phase is $\zeta_{x,y}=\arctan{z}/{z_{r_{x,y}}}$. The total power of this beam is $P$. 

 Here we set the parameters as follows: $w_{0_x}=\SI{1}{mm}$, $w_{0_y}=\SI{3}{mm}$, $P=1$, $\theta_s=30^{\circ}$ and $\lambda=\SI{1064}{nm}$. The propagation direction of the simple astigmatic Gaussian beam is set to $(0,0,1)$, and the beam center is positioned at $(x_{s_0},y_{s_0},z_{s_0})=(0.3,0.5,0)$. This beam can be decomposed into a superposition of a finite number of HG modes. The waist of these HG modes is $\SI{0.8}{mm}$, the direction of these HG modes is $(0,0,1)$ and the beam center is at $(0,0,0)$. The maximum mode order $N$ is set to $120$.
\begin{figure}[htbp]
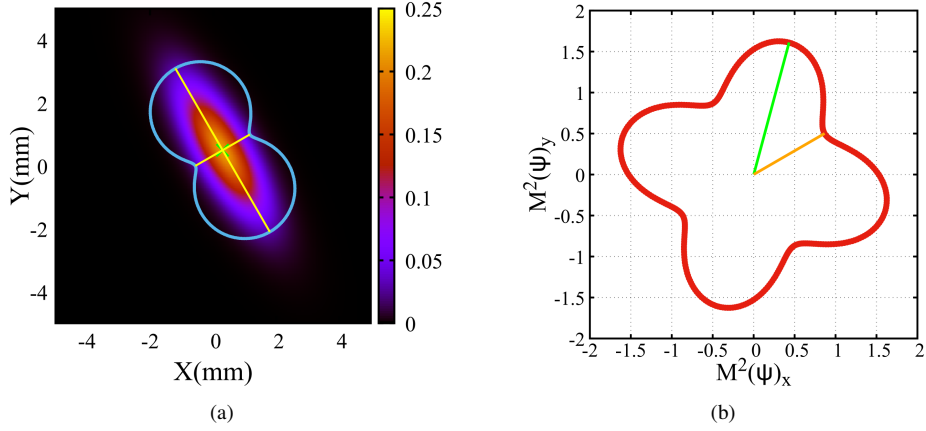

	\centering
	\subfigure[]{
	\begin{minipage}[t]{0.5\linewidth}
		\centering
		\includegraphics[width=0.9\textwidth]{mat_00_ang_30_tilt_0_inten_z_0.pdf}\label{intensity_sagb}
		
	\end{minipage}
	}
	\subfigure[]{
	\begin{minipage}[t]{0.46\linewidth}
		\centering
		\includegraphics[width=0.9\textwidth]{mat_00_ang_30_tilt_0_inten_z_0_m2.pdf}\label{sagb_m2}
		
	\end{minipage}
	}
	\centering
	\caption{(a). The intensity profile of the simple astigmatic Gaussian beam at $z= \SI{0}{mm}$. (b). The beam quality $M^2(\psi)$ factor curve of the simple astigmatic Gaussian beam for various angles $\psi$.}
\end{figure}

\Cref{intensity_sagb} illustrates the intensity profile of the simple astigmatic Gaussian beam at $z= \SI{0}{mm}$. The blue curve in \cref{intensity_sagb} delineates the boundary of the MSD spot sizes $w_{\psi}(\psi,z)$ for various rotation angles $\psi$ of this beam. Utilizing \cref{theta_ex1,theta_ex2}, we determine the spot rotation angles $\psi_{e1}$, $\psi_{e2}$ that correspond to the minimum and maximum MSD spot sizes, respectively. These angles are calculated as $\psi_{e1}=(30+5.831\times10^{-8})^{\circ}$ and $\psi_{e2}=(120+5.831\times10^{-8})^{\circ}$. By substituting these angles into \cref{wxhg_theta}, we calculate the minimum and maximum MSD spot sizes to be $w_{\psi}(\psi_{e1},0)= 1-\SI{1.542e-10 }{mm}$ and $w_{\psi}(\psi_{e2},0)= 3-\SI{2.444e-7}{mm}$ at $z=\SI{0}{mm}$, respectively. The yellow lines in \cref{intensity_sagb} represent these extreme spot sizes, demonstrating a good agreement with the theoretical results. The "green cross" symbol in \cref{intensity_sagb} marks the beam center, with its coordinates calculated by \cref{bahg_theta} as $(0.3+1.682\times10^{-8},0.5-2.977\times10^{-8},0)$. The direction, as calculated by \cref{propa_dir} is $(0,0,1)$. The beam center and direction are also found to be in good agreement with the theoretical results.

\Cref{sagb_m2} shows the beam quality $M^2(\psi)$ factor curve (red curve), as calculated by \cref{m2curve}, for various rotation angles $\psi$ of the simple astigmatic Gaussian beam. The orange and green lines in \cref{sagb_m2} represent the minimum and maximum beam quality $M^2(\psi)$ factors, respectively. These values are $M^2_{min}=1+2.366\times10^{-10}$ and $M^2_{max}=1.667$, with corresponding rotation angles $\psi_{min}=210^{\circ}$ and $\psi_{max}=75^{\circ}$. In this scenario, our results are in close alignment with the theoretical predictions.

\subsection{Oblique High-Order Hermite-Gaussian Beam}

The mode decomposition method is currently utilized across various domains, such as real-time measurement of beam quality \cite{2011OExpr..19.6741S} and rapid mode analysis of few-mode fibers \cite{2020NatCo..11.5507M}. In practical applications, there is often a discrepancy between the propagation directions of the incident beam and the basic HG modes for MEM, which may lead to the inaccurate calculation of the coefficients for each mode order. In this section,  we will assess the performance of our expressions under these conditions.

To simplify our analysis without sacrificing generality, we assume that the basic HG modes propagate in the $+z$ direction and the direction of the input oblique beam is $(0, a_{obl},1)$. We further assume that the input beam is an oblique high-order HG beam.

The complex amplitude of the oblique high-order HG beam is
 \begin{equation}
 \begin{split}
 	u_{h}=&\frac{c_{mn}}{w(z_l)}H_m\left(\frac{\sqrt{2}x_l}{w(z_l)}\right)H_n\left(\frac{\sqrt{2}y_l}{w(z_l)}\right)\exp{\left(-\frac{x_l^2+y_l^2}{w^2(z_l)}\right)}\\
	&\cdot\exp{\left(-ikz_l-ik\frac{x_l^2+y_l^2}{2R(z_l)}+i(m+n+1)\zeta(z_l)\right)},
	\label{hmn_complex_amp} 
\end{split}	
 \end{equation}
 where
 \begin{equation}
  \begin{split}
 x_l&=(\vec{x}-\vec{x_{h_0}})\cdot \vec{v_x},\ y_l=(\vec{x}-\vec{x_{h_0}})\cdot \vec{v_y},\ z_l=(\vec{x}-\vec{x_{h_0}})\cdot \vec{v_z},\\
\vec{v_y}&=(-\sqrt{1-(1+a_{o}^2)v_{y_y}^2},v_{y_y},-a_{o}\cdot v_{y_y}),\ v_{y_y}=\frac{\cos{(\theta_{h}+90^{\circ})}}{\sqrt{1+a_{o}^2}},\\
\vec{v_z}&=(0, \frac{a_{o}}{\sqrt{1+a_{o}^2}},\frac{1}{\sqrt{1+a_{o}^2}}),\ \vec{v_x}=\vec{v_y}\times\vec{v_z},\\
\vec{x}&=(x,y,z).
 \end{split}	
 \end{equation}
 
 We set the parameters as follows: $a_{o}=0.001$, $\theta_h=45^{\circ}$, $w_0=\SI{1}{mm}$, $P=1$, $\lambda=\SI{1064}{nm}$. The input beam center is located at $\vec{x_{h_0}}=(0.2,0.6,0)$, and the mode order $(m,n)$ of the input beam is $(3,1)$. This beam can be decomposed into a superposition of a finite number of HG modes. For the basic HG modes, the waist is $\SI{0.8}{mm}$, the beam center is at $(0,0,0)$, and the maximum mode order $N$ is set to $120$.
\begin{figure}[htbp]
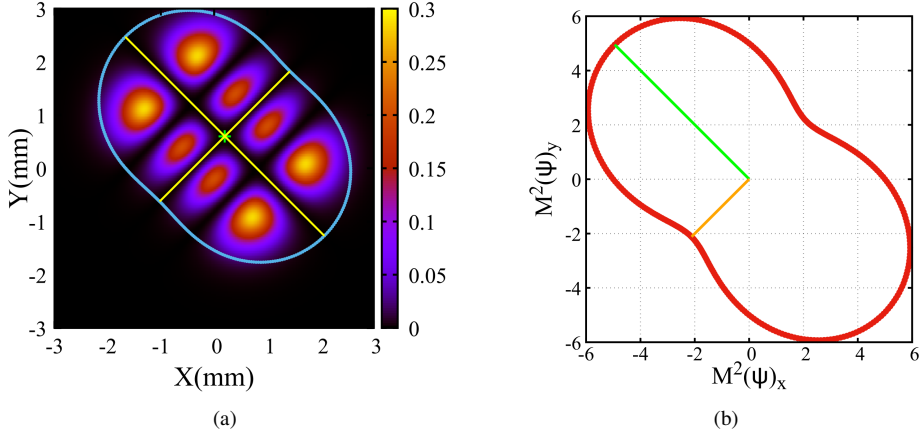

	\centering
	\subfigure[]{
	\begin{minipage}[t]{0.5\linewidth}
		\centering
		\includegraphics[width=0.9\textwidth]{mat_given_31_tilt_1e-3_120ord_w0_08_shift_mem_notilt_aper_ra_9_2_inten.pdf}\label{intensity_ohgb}
		
	\end{minipage}
	}
	\subfigure[]{
	\begin{minipage}[t]{0.46\linewidth}
		\centering
		\includegraphics[width=0.9\textwidth]{mat_given_31_tilt_1e-3_120ord_w0_08_shift_mem_notilt_aper_ra_9_2_ang_spot_m2.pdf}\label{ohgb_m2}
		
	\end{minipage}
	}
	\centering
	\caption{(a). The intensity profile of the oblique high-order HG beam at $z= \SI{0}{mm}$. (b).The beam quality $M^2(\psi)$ factor curve of the oblique high-order HG beam for various angles $\psi$.}
\end{figure}

\Cref{intensity_ohgb} illustrates the intensity profile of the oblique high-order HG beam at $z= \SI{0}{mm}$.  The blue curve in \cref{intensity_ohgb} delineates the boundary of the MSD spot sizes $w_{\psi}(\psi,z)$ for various rotation angles $\psi$ of this beam. Utilizing \cref{theta_ex1,theta_ex2}, we determine the spot rotation angles $\psi_{e1}$, $\psi_{e2}$ that correspond to the minimum and maximum MSD spot sizes, respectively. These angles are calculated as $\psi_{e1}=(45-3.581\times10^{-5})^{\circ}$ and $\psi_{e2}=(135-3.581\times10^{-5})^{\circ}$. By substituting these angles into \cref{wxhg_theta}, we calculate the minimum and maximum MSD spot sizes to be $w_{\psi}(\psi_{e1},0)=\SI{1.732}{mm}$ and $w_{\psi}(\psi_{e2},0)=\SI{2.646}{mm}$ at $z=\SI{0}{mm}$, respectively. The yellow lines in \cref{intensity_ohgb} represent these extreme spot sizes, demonstrating a good agreement with the theoretical predictions. The "green plus" symbol in \cref{intensity_ohgb} marks the beam center, with its coordinates calculated by \cref{bahg_theta} as $(0.2+1.882\times10^{-14},0.6-3.109\times10^{-15},0)$. The direction, as calculated by \cref{propa_dir} is $(0,1.0\times10^{-3}-1.0\times10^{-9}+1.508\times10^{-15},1-5.0\times10^{-7}+8.750\times10^{-13})$. The beam center and direction are also found to be in good agreement with the theoretical results.

\Cref{ohgb_m2} shows the beam quality $M^2(\psi)$ factor curve (red curve), as calculated by \cref{m2curve}, for various rotation angles $\psi$ of the oblique high-order HG Gaussian beam. The orange and green lines in \cref{ohgb_m2} represent the minimum and maximum beam quality $M^2(\psi)$ factors, respectively. These values are $M^2_{min}=3+5.840\times10^{-13}$ and $M^2_{max}=7+6.120\times10^{-13}$ with corresponding rotation angles $\psi_{min}=225^{\circ}$ and $\psi_{max}=135^{\circ}$. In this scenario, the results we obtained are in close agreement with the theoretical predictions. 

This implies that in practical applications, there is no necessity to meticulously align the propagation direction of each basic mode with that of the incident beam. In instances where the propagation direction of the basic HG modes does not align with that of the incident beam, employing this method enables us to determine the correct propagation direction of the incident beam. Subsequently, we can adjust the propagation direction of the basic HG modes to match this calculated direction. This adjustment allows us to accurately calculate the coefficients for each mode order.

\subsection{General Astigmatic Gaussian Beam}

In all the examples previously discussed, the spot of the incident beam does not rotate as it propagates. However, there are instances where the spot of the beam rotates during propagation, such as with GAGB. In this subsection, we will assess the effectiveness of our expressions in this situation. 

When a fundamental Gaussian beam traverses a nonorthogonal optical system, it transforms into a GAGB. The complex amplitude of the input GAGB is given by \cite{kochkina2013stigmatic}
\begin{equation}
\begin{split}
u_{g}=&E_{0_g}\exp{\left(-i\phi_{ac}+i\eta-i\frac{k}{2}\textbf{r}^{T}\textbf{Q}\textbf{r}\right)},
\end{split}
\end{equation}
where the tensor $\textbf{Q}$ is defined as:
\begin{equation}
\textbf{Q}=
\begin{pmatrix}
\frac{\cos^2\theta_g}{q_1}+\frac{\sin^2\theta_g}{q_2} & \frac{1}{2}\sin{2\theta_g}\left(\frac{1}{q_1}-\frac{1}{q_2}\right)\\
\frac{1}{2}\sin{2\theta_g}\left(\frac{1}{q_1}-\frac{1}{q_2}\right) & \frac{\sin^2\theta_g}{q_1}+\frac{\cos^2\theta_g}{q_2} 
\end{pmatrix}.
\end{equation}
Here, $q_1$, $q_2$ are the q-parameters, $\theta_g$ is the complex angle, $\textbf{r}=(x,y,z)$ represents the position vector, $\phi_{ac}$ is the accumulated optical path length, $\eta=\frac{1}{2}\left(\arctan{{Re(q_1)}/{Im(q_1)}}+\arctan{{Re(q_2)}/{Im(q_2)}}\right)$ is the Gouy phase, and $E_{0_g}=\sqrt{{P}\sqrt{4Im(Q_{11})Im(Q_{22})-Im(Q_{12}+Q_{21})^2}/{\lambda}}$, with $Q_{ij}$ denoting the $(i,j)$-th element of the tensor $\textbf{Q}$.

In this case, we assume the following paprameters: $q_1=66i$, $q_2=500+266i$, $\theta_g={\pi}/{9}+({\pi}/{18})i$, $\lambda=\SI{1064}{nm}$, and $\phi_{ac}=0$. The beam center of the GAGB is positioned at $(0,0,0)$, and the beam direction is aligned with $(0,0,1)$. This beam can be decomposed into a superposition of a finite number of HG modes. The waist of the basic HG modes is $\SI{0.2}{mm}$, with these modes propagating in the $(0,0,1)$ direction and having their beam center at $(0,0,0)$. The maximum mode order $N$ is set to $120$.
\begin{figure}[htbp]
	\centering
	\subfigure[]{
	\begin{minipage}[t]{0.327\linewidth}
		\centering
		\includegraphics[width=0.9\textwidth]{mat_gagb_ra_3_w0_02_ord_120_theta_20_10_q2_266i_2_inten_z_0.pdf}\label{intensity_gagb_z_0}
		
	\end{minipage}
	}
	\subfigure[]{
	\begin{minipage}[t]{0.32\linewidth}
		\centering
		\includegraphics[width=0.9\textwidth]{mat_gagb_ra_3_w0_02_ord_120_theta_20_10_q2_266i_2_inten_z_100.pdf}\label{intensity_gagb_z_100}
		
	\end{minipage}
	
	}
		\subfigure[]{
	\begin{minipage}[t]{0.3\linewidth}
		\centering
		\includegraphics[width=0.9\textwidth]{mat_gagb_ra_3_w0_02_ord_120_theta_20_10_q2_266i_2_ang_spot_m2.pdf}\label{m2_gagb}
		
	\end{minipage}
	}
		\subfigure[]{
	\begin{minipage}[t]{0.313\linewidth}
		\centering
		\includegraphics[width=0.9\textwidth]{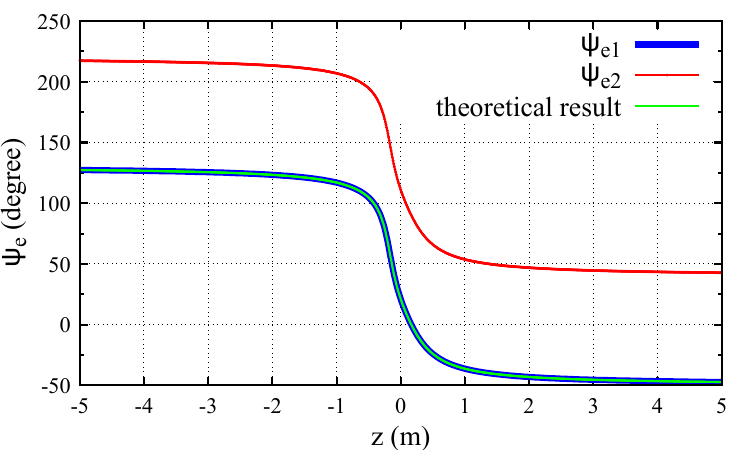}\label{ang_z_gagb}
		
	\end{minipage}
	}
		\subfigure[]{
	\begin{minipage}[t]{0.313\linewidth}
		\centering
		\includegraphics[width=0.9\textwidth]{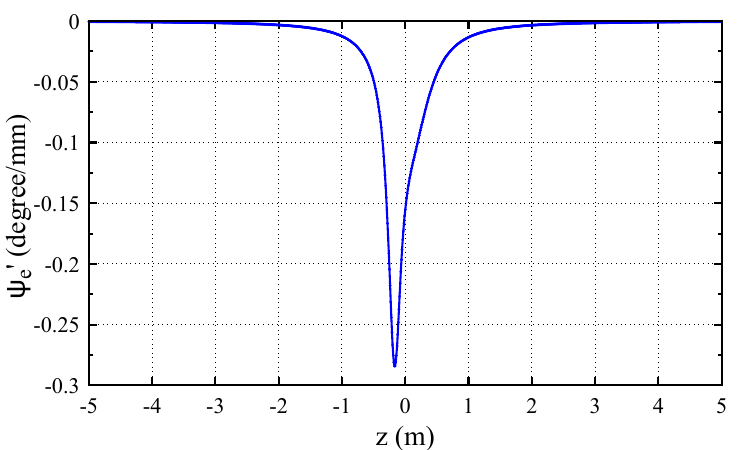}\label{ang_velo_z_gagb}
		
	\end{minipage}
	}
	\subfigure[]{
	\begin{minipage}[t]{0.313\linewidth}
		\centering
		\includegraphics[width=0.9\textwidth]{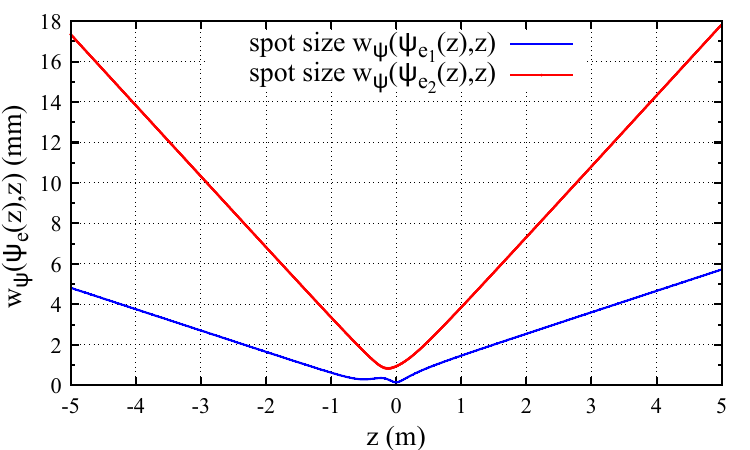}\label{spot_z_gagb}
		
	\end{minipage}
	}

	\centering
	\caption{(a). The intensity profile of the GAGB at $z= \SI{0}{mm}$. (b). The intensity profile of the GAGB at $z= \SI{100}{mm}$. (c). The beam quality $M^2(\psi)$ factor curve of the GAGB for various angles $\psi$. (d). The spot rotation angles $\psi_{e1}$, $\psi_{e2}$ correspond to different values of $z$. (e). The corresponding spot rotation angular speed $\psi_{e}'$ for various values of $z$. (f). The corresponding extreme spot sizes for the spot rotation angles $\psi_{e1}$, $\psi_{e2}$ at various values of $z$.}
\end{figure}

\Cref{intensity_gagb_z_0} and \cref{intensity_gagb_z_100} illustrate the intensity profiles of the GAGB at $z= \SI{0}{mm}$ and $z= \SI{100}{mm}$, respectively. The blue curves in \cref{intensity_gagb_z_0} and \cref{intensity_gagb_z_100} outline the boundary of the MSD spot sizes $w_{\psi}(\psi,z)$ for various rotation angles $\psi$ of this beam. Utilizing \cref{theta_ex1,theta_ex2}, we determine the spot rotation angles $\psi_{e1}$, $\psi_{e2}$ that correspond to the minimum and maximum MSD spot sizes, respectively. These angles are calculated as $\psi_{e1}=21.046^{\circ}$, $\psi_{e2}=111.046^{\circ}$ for $z= \SI{0}{mm}$ and $\psi_{e1}=7.417^{\circ}$, $\psi_{e2}=97.417^{\circ}$ for $z= \SI{100}{mm}$. By substituting these angles into \cref{wxhg_theta}, we calculate the minimum and maximum MSD spot sizes. For $z= \SI{0}{mm}$, the extreme spot sizes are $w_{\psi}(\psi_{e1},0) =\SI{0.147}{mm}$, $w_{\psi}(\psi_{e2},0)=\SI{0.940}{mm}$. For $z= \SI{100}{mm}$, the extreme spot sizes are $w_{\psi}(\psi_{e1},100)=\SI{0.262}{mm}$, $w_{\psi}(\psi_{e2},100)=\SI{1.113}{mm}$. The yellow lines in \cref{intensity_gagb_z_0} and \cref{intensity_gagb_z_100} represent these extreme spot sizes. The "green cross" symbols in \cref{intensity_gagb_z_0} and \cref{intensity_gagb_z_100} mark the beam centers, with its coordinates calculated by \cref{bahg_theta} as $(0,0,0)$ and $(0,0,100)$. The direction, as calculated by \cref{propa_dir}, is $(0,0,1)$. The beam centers and direction are also found to be in good agreement with the theoretical results.

\Cref{m2_gagb} presents the beam quality $M^2(\psi)$ factor curve (red curve), as calculated by \cref{m2curve}, for various rotation angles $\psi$ of the GAGB. The orange and green lines in \cref{m2_gagb} represent the minimum and maximum beam quality $M^2(\psi)$ factors, respectively. These values are $M^2_{min}=1.449$ and $M^2_{max}=3.283$, with corresponding rotation angles $\psi_{min}=200^{\circ}$ and $\psi_{max}=65^{\circ}$. 

\Cref{ang_z_gagb} shows the variations of the spot rotation angles $\psi_{e1}$ and $\psi_{e2}$ as functions of $z$. The blue line represents $\psi_{e1}$, the red line represents $\psi_{e2}$, and the green line represents the theoretical result. The variations in the spot rotation angular speed $\psi_{e}'$ for different values of $z$ are depicted in \cref{ang_velo_z_gagb}. The minimum spot rotation angular speed, $\psi_{e_{min}}'=-0.284^{\circ}/\SI{}{mm}$, occurs at $z=\SI{-167.3}{mm}$. \Cref{spot_z_gagb} displays the corresponding extreme spot sizes for $\psi_{e1}$ and $\psi_{e2}$ at various values of $z$. 

The spot rotation angle and the spot rotation angular speed of the GAGB, as calculated using our derived formulas, are in excellent agreement with both the observed beam intensity patterns and the theoretical results. The fact that the spot rotation angular speed is always negative confirms that the GAGB maintains a clockwise rotation as $z$ increases. As $z$ transitions from negative to positive infinity, the rotation angle of the GAGB's spot experiences a 180-degree shift.  Consistent with our theoretical predictions, the spot rotation angles $\psi_{e1}$ and $\psi_{e2}$ indeed converge as $z$ approaches either positive or negative infinity.

\subsection{elliptical vortex Hermite-Gaussian beam}

To date, the study of vortex laser beams has garnered considerable attention in both experimental and theoretical realms, due to their broad applicative potential \cite{2024OQEle..56.1356A}. In recent years, numerous models have been proposed to describe this class of beams, with thorough investigations into their propagation behaviors across various media \cite{2011OptL...36.1617D,2015OptL...40..701K}. Furthermore, significant research efforts have been directed toward investigating the beam quality $M^2$ factors of various vortex beams \cite{2010OptLT..42..489Z,2018OptLT..99..191M}. In this subsection, we will examine the performance of our formulas using the vortex Hermite-Gaussian beam as a representative case study.

The complex amplitude of the vortex Hermite-Gaussian(vHG) beam is given by \cite{2015OptL...40..701K}
\begin{equation}
   u_n(x,y,z=0)=i^n\exp{\left(-\frac{x^2}{p^2}-\frac{y^2}{q^2}\right)}(1+a^2)^{-\frac{n}{2}}\sum_{j=0}^{n}\frac{n!(ia)^j}{j!(n-j)!}H_j(\frac{x}{c})H_{n-j}(\frac{y}{d}),
\end{equation}
where $p=\sqrt{2}c=w_{x0}$, $q=\sqrt{2}d=w_{y0}$.The normalized OAM, defined as the OAM projection onto the optical axis divided by the beam power, is given for the vHG beam by \cite{2015OptL...40..701K}
\begin{equation}
\frac{J_z}{P}=\frac{-na}{1+a^2}\frac{w_{x0}^2+w_{y0}^2}{w_{x0}w_{y0}}.
\end{equation}

We test the performance of our formulas under two distinct scenarios: 1. vHG modes where $w_{y0}=w_{x0}$; 2. elliptical vHG beams where $w_{y0}\neq w_{x0}$. The configuration parameters for each scenario are detailed in \cref{vhg_para}. Notably, we have specifically set the normalized OAM of the two beams to be identical.

\begin{table}[htbp]
\centering
\caption{\bf Configuration parameters for the vHG beam and elliptical vHG beam }
\begin{tabular}{ccccccc}
\hline
 & n & a & $w_{x0}$ & $w_{y0}$ &$\lambda$ & $\frac{J_z}{P}$ \\
\hline
vHG mode & 3 & 0.5 & $10\lambda$ & $10\lambda$ & $\SI{532}{nm}$ & -2.4\\
elliptical vHG beam& 3 & 0.25569 &$10\lambda$ & $30\lambda$ & $\SI{532}{nm}$ & -2.4 \\
\hline
\end{tabular}
  \label{vhg_para}
\end{table}

 The beam centers for both the vHG mode and the elliptical vHG beam are positioned at $(0,0,0)$, and their beam directions are aligned with $(0,0,1)$. These beams can be decomposed into a superposition of a finite number of HG modes. The waist of these basic HG modes is $\SI{0.005}{mm}$, their direction is $(0,0,1)$, and the beam center is $(0,0,0)$. The maximum mode order $N$ is set to $120$.

\begin{figure}[htbp]
	\centering
	\subfigure[]{
	\begin{minipage}[t]{0.32\linewidth}
		\centering
		\includegraphics[width=0.9\textwidth]{mat_vhg_wy_wx_n_3_ord_120_w0_0005_z_0_inten_z_0.pdf}\label{intensity_vhg_wywx_z_0}
		
	\end{minipage}
	}
	\subfigure[]{
	\begin{minipage}[t]{0.33\linewidth}
		\centering
		\includegraphics[width=0.9\textwidth]{mat_vhg_wy_wx_n_3_ord_120_w0_0005_z_0_inten_z_10.pdf}\label{intensity_vhg_wywx_z_10}
		
	\end{minipage}
	
	}
		\subfigure[]{
	\begin{minipage}[t]{0.29\linewidth}
		\centering
		\includegraphics[width=0.9\textwidth]{mat_vhg_wy_wx_n_3_ord_120_w0_0005_z_0_ang_spot_m2.pdf}\label{m2_vhg_wywx}
		
	\end{minipage}
	}
		\subfigure[]{
	\begin{minipage}[t]{0.313\linewidth}
		\centering
		\includegraphics[width=0.9\textwidth]{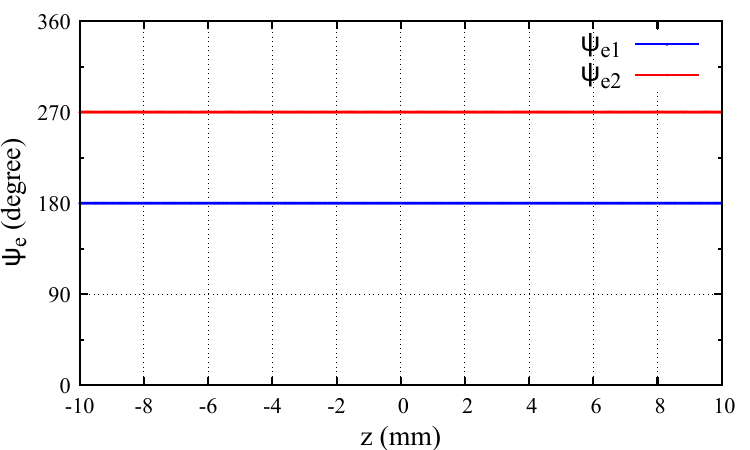}\label{ang_z_vhg_wywx}
		
	\end{minipage}
	}
		\subfigure[]{
	\begin{minipage}[t]{0.313\linewidth}
		\centering
		\includegraphics[width=0.9\textwidth]{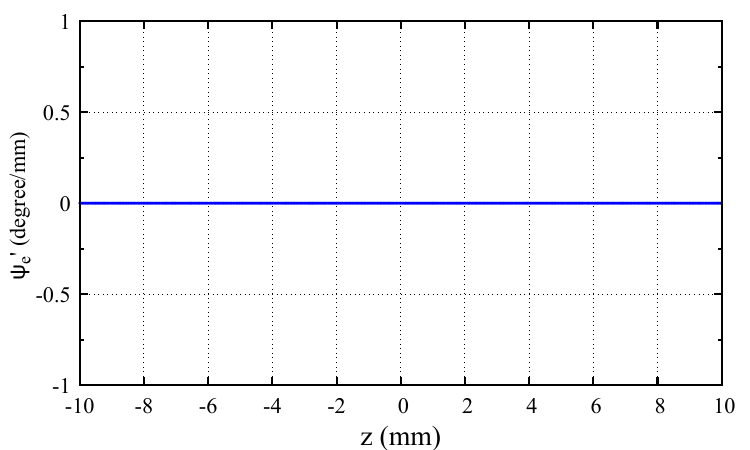}\label{ang_velo_z_vhg_wywx}
		
	\end{minipage}
	}
		\subfigure[]{
	\begin{minipage}[t]{0.313\linewidth}
		\centering
		\includegraphics[width=0.9\textwidth]{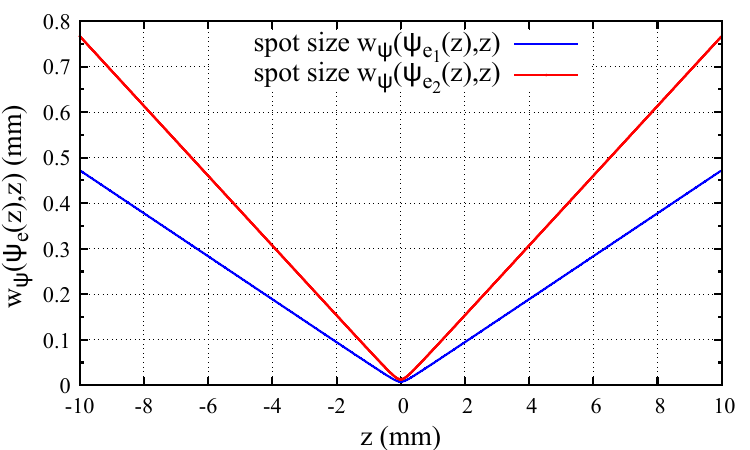}\label{spot_z_vhg_wywx}
		
	\end{minipage}
	}
	\centering
	\caption{(a). The intensity profile of the vHG mode ($w_{y0}=w_{x0}$) at $z= \SI{0}{mm}$. (b). The intensity profile of the vHG mode at $z= \SI{10}{mm}$. (c). The beam quality $M^2(\psi)$ factor curve of the vHG mode for various angles $\psi$. (d). The spot rotation angles $\psi_{e1}$, $\psi_{e2}$ correspond to different values of $z$. (e). The corresponding spot rotation angular speed $\psi_{e}'$ for various values of $z$. (f).The corresponding extreme spot sizes for the spot rotation angles $\psi_{e1}$, $\psi_{e2}$ at various values of $z$.}\label{vhgwywx}
\end{figure}

\Cref{intensity_vhg_wywx_z_0} and \cref{intensity_vhg_wywx_z_10} illustrate the intensity profiles of the vHG mode ($w_{y0}=w_{x0}$) at $z= \SI{0}{mm}$ and $z= \SI{10}{mm}$, respectively. The blue curves in \cref{intensity_vhg_wywx_z_0} and \cref{intensity_vhg_wywx_z_10} delineate the boundary of the MSD spot sizes $w_{\psi}(\psi,z)$ for various rotation angles $\psi$ of this beam. By utilizing \cref{theta_ex1,theta_ex2}, we determine the spot rotation angles $\psi_{e1}$, $\psi_{e2}$ that correspond to the minimum and maximum MSD spot sizes, respectively. These angles are calculated as $\psi_{e1}=180^{\circ}$, $\psi_{e2}=270^{\circ}$ for $z= \SI{0}{mm}$ and $\psi_{e1}=180^{\circ}$, $\psi_{e2}=270^{\circ}$ for $z= \SI{10}{mm}$. Upon substituting these angles into \cref{wxhg_theta}, we calculate the minimum and maximum MSD spot sizes. For $z= \SI{0}{mm}$, the extreme spot sizes are $w_{\psi}(\psi_{e1},0)= \SI{0.008}{mm}$, $w_{\psi}(\psi_{e2},0)=\SI{0.013}{mm}$. For $z= \SI{10}{mm}$, the extreme spot sizes are $w_{\psi}(\psi_{e1},10)=\SI{0.472}{mm}$, $w_{\psi}(\psi_{e2},10)=\SI{0.767}{mm}$. The yellow lines in \cref{intensity_vhg_wywx_z_0} and \cref{intensity_vhg_wywx_z_10} represent these extreme spot sizes. The "green cross" symbols in \cref{intensity_vhg_wywx_z_0} and \cref{intensity_vhg_wywx_z_10} mark the beam centers, with its coordinates calculated by \cref{bahg_theta} as $(0,0,0)$ and $(0,0,10)$. The direction, as calculated by \cref{propa_dir} is $(0,0,1)$. 

\Cref{m2_vhg_wywx} presents the beam quality $M^2(\psi)$ factor curve (red curve), as calculated by \cref{m2curve}, for various rotation angles $\psi$ of the vHG mode. The orange and green lines in \cref{m2_vhg_wywx} represent the minimum and maximum beam quality $M^2(\psi)$ factors, respectively. These values are $M^2_{min}=2.200$ and $M^2_{max}=5.800$, with corresponding rotation angles $\psi_{min}=0^{\circ}$ and $\psi_{max}=90^{\circ}$. 

\Cref{ang_z_vhg_wywx} shows the variations of the spot rotation angles $\psi_{e1}$ and $\psi_{e2}$ as functions of $z$. The blue line represents $\psi_{e1}$, and the red line represents $\psi_{e2}$. The variations in the spot rotation angular speed $\psi_{e}'$ for different values of $z$ are depicted in \cref{ang_velo_z_vhg_wywx}.  \Cref{spot_z_vhg_wywx} displays the corresponding extreme spot sizes of $\psi_{e1}$ and $\psi_{e2}$ at various values of $z$. The vHG mode maintains its shape upon propagation, and the spot rotation angular speed remains zero throughout the propagation.

\begin{figure}[htbp]
	\centering
	\subfigure[]{
	\begin{minipage}[t]{0.325\linewidth}
		\centering
		\includegraphics[width=0.9\textwidth]{mat_vhg_wy_3wx_n_3_ord_120_w0_0005_z_0_a_025569_inten_z_0.pdf}\label{intensity_vhg_wy3wx_z_0_a025}
		
	\end{minipage}
	}
	\subfigure[]{
	\begin{minipage}[t]{0.325\linewidth}
		\centering
		\includegraphics[width=0.9\textwidth]{mat_vhg_wy_3wx_n_3_ord_120_w0_0005_z_0_a_025569_inten_z_1.pdf}\label{intensity_vhg_wy3wx_z_1_a025}
		
	\end{minipage}
	
	}
		\subfigure[]{
	\begin{minipage}[t]{0.29\linewidth}
		\centering
		\includegraphics[width=0.9\textwidth]{mat_vhg_wy_3wx_n_3_ord_120_w0_0005_z_0_a_025569_ang_spot_m2_z.pdf}\label{m2_vhg_wy3wx_a025}
		
	\end{minipage}
	}
		\subfigure[]{
	\begin{minipage}[t]{0.313\linewidth}
		\centering
		\includegraphics[width=0.9\textwidth]{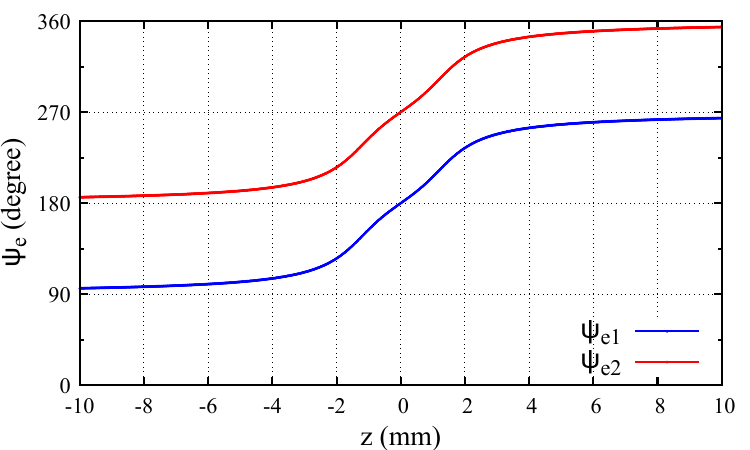}\label{ang_z_vhg_wy3wx_a025}
		
	\end{minipage}
	}
		\subfigure[]{
	\begin{minipage}[t]{0.313\linewidth}
		\centering
		\includegraphics[width=0.9\textwidth]{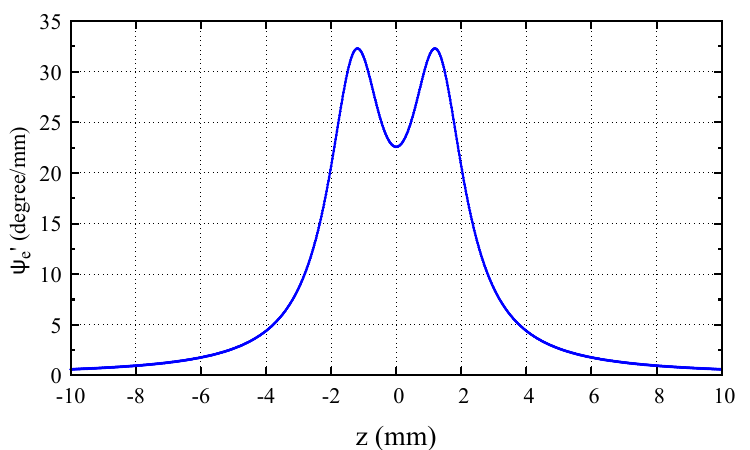}\label{ang_velo_z_vhg_wy3wx_a025}
		
	\end{minipage}
	}
	\subfigure[]{
	\begin{minipage}[t]{0.313\linewidth}
		\centering
		\includegraphics[width=0.9\textwidth]{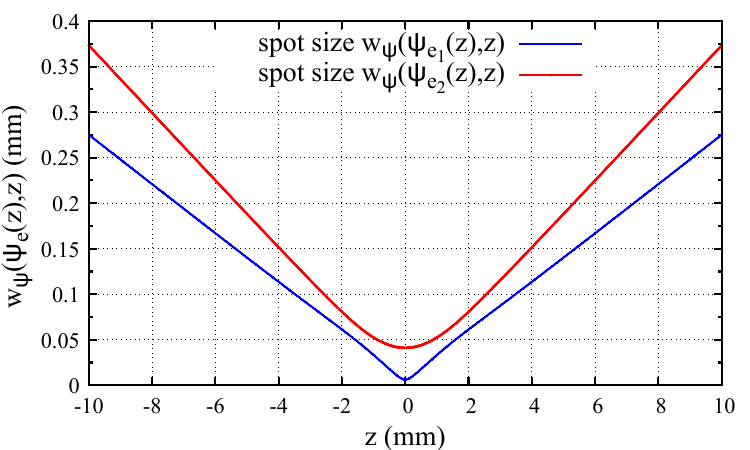}\label{spot_z_vhg_wy3wx_a025}
		
	\end{minipage}
	}
	\centering
	\caption{(a). The intensity profile of the elliptical vHG beam ($w_{y0}=3w_{x0}$) at $z= \SI{0}{mm}$. (b). The intensity profile of the elliptical vHG beam at $z= \SI{1}{mm}$. (c). The beam quality $M^2(\psi)$ factor curve of the elliptical vHG beam for various angles $\psi$. (d). The spot rotation angles $\psi_{e1}$, $\psi_{e2}$ correspond to different values of $z$. (e). The corresponding spot rotation angular speed $\psi_{e}'$ for various values of $z$. (f). The corresponding extreme spot sizes for the spot rotation angles $\psi_{e1}$, $\psi_{e2}$ at various values of $z$.}\label{vhgwy3wx}
\end{figure}

\Cref{intensity_vhg_wy3wx_z_0_a025} and \cref{intensity_vhg_wy3wx_z_1_a025} illustrate the intensity profiles of the elliptical vHG beam ($w_{y0}=3w_{x0}$) at $z= \SI{0}{mm}$ and $z= \SI{1}{mm}$, respectively. The blue curves in \cref{intensity_vhg_wy3wx_z_0_a025} and \cref{intensity_vhg_wy3wx_z_1_a025} delineate the boundary of the MSD spot sizes $w_{\psi}(\psi,z)$ for various rotation angles $\psi$ of this beam. Utilizing \cref{theta_ex1,theta_ex2}, we determine the spot rotation angles $\psi_{e1}$, $\psi_{e2}$  that correspond to the minimum and maximum MSD spot sizes, respectively. These angles are calculated as $\psi_{e1}=180^{\circ}$, $\psi_{e2}=270^{\circ}$ for $z= \SI{0}{mm}$ and $\psi_{e1}=26.071^{\circ}$, $\psi_{e2}=116.071^{\circ}$  for $z= \SI{1}{mm}$. By substituting these angles into \cref{wxhg_theta}, we calculate the minimum and maximum MSD spot sizes. For $z= \SI{0}{mm}$, the extreme spot sizes are $w_{\psi}(\psi_{e1},0)=\SI{0.006}{mm}$, $w_{\psi}(\psi_{e2},0)=\SI{0.041}{mm}$. For $z= \SI{1}{mm}$, the extreme spot sizes are $w_{\psi}(\psi_{e1},1)=\SI{0.033}{mm}$, $w_{\psi}(\psi_{e2},1)=\SI{0.052}{mm}$. The yellow lines in \cref{intensity_vhg_wy3wx_z_0_a025} and \cref{intensity_vhg_wy3wx_z_1_a025} represent these extreme spot sizes. The "green cross" symbols in \cref{intensity_vhg_wy3wx_z_0_a025} and \cref{intensity_vhg_wy3wx_z_1_a025} mark the beam centers, with its coordinates calculated by \cref{bahg_theta} as $(0,0,0)$ and $(0,0,1)$. The direction, as calculated by \cref{propa_dir}, is $(0,0,1)$. 

\Cref{m2_vhg_wy3wx_a025} presents the beam quality $M^2(\psi)$ factor curve (red curve), as calculated by \cref{m2curve}, for various rotation angles $\psi$ of the elliptical vHG beam. The orange and green lines in \cref{m2_vhg_wy3wx_a025} represent the minimum and maximum beam quality $M^2(\psi)$ factors, respectively. These values are $M^2_{min}=1.368$ and $M^2_{max}=6.632$, with corresponding rotation angles $\psi_{min}=0^{\circ}$ and $\psi_{max}=90^{\circ}$. 

\Cref{ang_z_vhg_wy3wx_a025} shows the variations of the spot rotation angles $\psi_{e1}$ and $\psi_{e2}$ as functions of $z$. The blue line represents $\psi_{e1}$, and the red line represents $\psi_{e2}$. The variations in the spot rotation angular speed $\psi_{e}'$ for different values of $z$ are depicted in \cref{ang_velo_z_vhg_wy3wx_a025}. The maximum spot rotation angular speed, $\psi_{e_{max}}'=32.286^{\circ}/\SI{}{mm}$, occurs at $z=\SI{-1.18}{mm}$ and $z=\SI{1.18}{mm}$. The minimal spot rotation angular speed, $\psi_{e_{min}}'=22.573^{\circ}/\SI{}{mm}$, occurs at $z=\SI{0}{mm}$. \Cref{spot_z_vhg_wy3wx_a025} displays the corresponding extreme spot sizes for $\psi_{e1}$ and $\psi_{e2}$ at various values of $z$. 

As theory predicts, although the vHG mode and the elliptical vHG beam discussed in this subsection possess the same orbital angular momentum, the spot of the former does not rotate with $z$, whereas the spot of the latter does rotate with $z$. We first show the spot rotation angle and spot rotation angular speed for the elliptical vHG beam in this subsection. The spot rotation angle and the spot rotation angular speed, as calculated by our equations, are in good agreement with the observed beam intensity patterns. Our formula can efficiently distinguish between VOAM beams and AOAM beams. The spot rotation angular speed of the elliptical vHG beam is always positive, indicating that this beam maintains a counterclockwise rotation as $z$ increases. Consistent with our theoretical analysis from the previous section, the spot rotation angles $\psi_{e1}$ and $\psi_{e2}$ indeed converge as $z$ approaches either positive or negative infinity.

\section{Conclusion}
In this paper, we introduce rapid calculation formulas for key parameters of the general MEM beam, including direction, divergence angle, waist position, waist radius, and the traditional beam quality $M^2$ factor. Additionally, we propose new definitions for assessing beam quality: the $M^2(\psi)$ factor, the spot rotation angle $\psi_{e}$, and the spot rotation angular speed $\psi_{e}'$.
We also define the beam center $s_{\psi}(\psi,z)$, spot size $w_{\psi}(\psi,z)$, waist position, waist radius, and divergence angle for a general beam along a direction that forms an angle $\psi$ with the $x$-axis in the plane perpendicular to the $z$-axis, providing rapid calculation formulas for these parameters as well.
For the first time, this paper offers a systematic approach to characterizing the rotational properties of general beams. It is proved that in any given detection plane, there are only two extreme spot sizes, with the angle between the maximum and minimum spot angles consistently being $90^{\circ}$ during propagation. Furthermore, we demonstrate that the spot rotation angles converge as $z$ approaches positive or negative infinity.

Subsequently, we evaluate the performance of our formulas across various scenarios, yielding highly satisfactory results. Our equations accurately predict the spot size, beam center, direction, and beam quality $M^2(\psi)$ factor for both the rotated simple astigmatic Gaussian beam and the oblique high-order HG beam, aligning perfectly with theoretical predictions. This accuracy eliminates the need for meticulous alignment of each basic mode's propagation direction with the incident beam in practical applications.
When the propagation direction of the basic HG modes deviates from that of the incident beam, our method accurately calculates the incident beam's correct propagation direction. By adjusting the basic HG modes to match this direction, we can precisely determine the coefficients for each mode order.
We also assess our formulas in rotational beams, such as the GAGB and the elliptical vHG beam. The calculated spot rotation angle and angular speed for the GAGB closely match observed beam intensity patterns and theoretical results. Similarly, for the elliptical vHG beam, our equations show good agreement with observed patterns.
Our formulas effectively differentiate between VOAM and AOAM beams. As predicted, the spot rotation angles $\psi_{e1}$ and $\psi_{e2}$ converge as $z$ approaches positive or negative infinity, and the angle between the maximum and minimum spot angles remains consistently $90^{\circ}$ during propagation in both GAGB and elliptical vHG beam scenarios.
Furthermore, we present the beam quality $M^2(\psi)$ factor curve and the extreme $M^2(\psi)$ factors for these four beams, demonstrating the practical application benefits of our research for vortex beams.

\section{Acknowledgements}
 This work has been supported in part by the National Key Research and Development Program of China under Grant No.2020YFC2201501, the National Science Foundation of China (NSFC) under Grants No. 12147103 (special fund to the center for quanta-to-cosmos theoretical physics), No. 11821505, the Strategic Priority Research Program of the Chinese Academy of Sciences under Grant No. XDB23030100. In this paper, the part of the numerical computation is finished by TAIJI Cluster.
 
\appendix
\section{Coefficients}\label{appen_coe}
\begin{equation}
\begin{split}
&a_x=-4 \left(\sum_{m=0}^{N}\sum_{n=0}^{N-m}|a_{mn}||a_{(m+1)n}|\sqrt{m+1}\sin{(\beta_{mn}-\beta_{(m+1)n})}\right)^2\\
&+\sum_{m=0}^{N}\sum_{n=0}^{N-m}|a_{mn}|^2(2m+1)P_{MEM}\\
&-2P_{MEM} \sum_{m=0}^{N}\sum_{n=0}^{N-m}|a_{mn}||a_{(m+2)n}|\cos{(\beta_{mn}-\beta_{(m+2)n})}\sqrt{(m+2)(m+1)}.
\end{split}
\end{equation}

\begin{small}
\begin{equation}
\begin{split}
&b_x=4P_{MEM}\sum_{m=0}^{N}\sum_{n=0}^{N-m}|a_{mn}||a_{(m+2)n}|\sin{(\beta_{mn}-\beta_{(m+2)n})}\sqrt{(m+2)(m+1)}\\
&-8\sum_{m=0}^{N}\sum_{n=0}^{N-m}\sum_{l=0}^{N}\sum_{h=0}^{N-l}|a_{mn}||a_{(m+1)n}||a_{lh}||a_{(l+1)h}|\sqrt{m+1}\sqrt{l+1}\sin{(\beta_{mn}-\beta_{(m+1)n})}\cos{(\beta_{lh}-\beta_{(l+1)h})}.
\end{split}
\end{equation}
\end{small}

\begin{equation}
\begin{split}
&c_x=-4 \left(\sum_{m=0}^{N}\sum_{n=0}^{N-m}|a_{mn}||a_{(m+1)n}|\sqrt{m+1}\cos{(\beta_{mn}-\beta_{(m+1)n})}\right)^2\\
&+\sum_{m=0}^{N}\sum_{n=0}^{N-m}|a_{mn}|^2(2m+1)P_{MEM} \\
&+2P_{MEM} \sum_{m=0}^{N}\sum_{n=0}^{N-m}|a_{mn}||a_{(m+2)n}|\cos{(\beta_{mn}-\beta_{(m+2)n})}\sqrt{(m+2)(m+1)}.
\end{split}
\end{equation}

\begin{equation}
\begin{split}
&a_y=-4 \left(\sum_{m=0}^{N}\sum_{n=0}^{N-m}|a_{mn}||a_{m(n+1)}|\sqrt{n+1}\sin{(\beta_{mn}-\beta_{m(n+1)})}\right)^2\\
&+\sum_{m=0}^{N}\sum_{n=0}^{N-m}|a_{mn}|^2(2n+1)P_{MEM}\\
&-2P_{MEM} \sum_{m=0}^{N}\sum_{n=0}^{N-m}|a_{mn}||a_{m(n+2)}|\cos{(\beta_{mn}-\beta_{m(n+2)})}\sqrt{(n+2)(n+1)}.
\end{split}
\end{equation}

\begin{small}
\begin{equation}
\begin{split}
&b_y=4P_{MEM}\sum_{m=0}^{N}\sum_{n=0}^{N-m}|a_{mn}||a_{m(n+2)}|\sin{(\beta_{mn}-\beta_{m(n+2)})}\sqrt{(n+2)(n+1)}\\
&-8\sum_{m=0}^{N}\sum_{n=0}^{N-m}\sum_{l=0}^{N}\sum_{h=0}^{N-l}|a_{mn}||a_{m(n+1)}||a_{lh}||a_{l(h+1)}|\sqrt{n+1}\sqrt{h+1}\sin{(\beta_{mn}-\beta_{m(n+1)})}\cos{(\beta_{lh}-\beta_{l(h+1)})}.
\end{split}
\end{equation}
\end{small}

\begin{equation}
\begin{split}
&c_y=-4 \left(\sum_{m=0}^{N}\sum_{n=0}^{N-m}|a_{mn}||a_{m(n+1)}|\sqrt{n+1}\cos{(\beta_{mn}-\beta_{m(n+1)})}\right)^2\\
&+\sum_{m=0}^{N}\sum_{n=0}^{N-m}|a_{mn}|^2(2n+1)P_{MEM}\\
&+2P_{MEM} \sum_{m=0}^{N}\sum_{n=0}^{N-m}|a_{mn}||a_{m(n+2)}|\cos{(\beta_{mn}-\beta_{m(n+2)})}\sqrt{(n+2)(n+1)}.
\end{split}
\end{equation}

\begin{equation}
\begin{split}
   \mathcal W_1=&\sum_{m=0}^{N}\sum_{n=0}^{N-m}|a_{mn}||a_{(m+1)(n+1)}|\cos{(\beta_{mn}-\beta_{(m+1)(n+1)}-2\zeta(z))}\sqrt{(m+1)(n+1)}\\
  &+\sum_{m=0}^{N}\sum_{n=0}^{N-m}|a_{m(n+1)}||a_{(m+1)n}|\cos{(\beta_{m(n+1)}-\beta_{(m+1)n})}\sqrt{(m+1)(n+1)}\label{w1}.
\end{split}
\end{equation}

\begin{footnotesize}
\begin{equation}
\begin{split}
   \mathcal W_1'=&\frac{d\mathcal W_1}{dz}=2\frac{d\zeta(z)}{dz}\sum_{m=0}^{N}\sum_{n=0}^{N-m}|a_{mn}||a_{(m+1)(n+1)}|\sin{(\beta_{mn}-\beta_{(m+1)(n+1)}-2\zeta(z))}\sqrt{(m+1)(n+1)}\\
   =&2\frac{z_r}{z^2+z_r^2}\sum_{m=0}^{N}\sum_{n=0}^{N-m}|a_{mn}||a_{(m+1)(n+1)}|\sin{(\beta_{mn}-\beta_{(m+1)(n+1)}-2\zeta(z))}\sqrt{(m+1)(n+1)}.
 \end{split}
\end{equation}
\end{footnotesize}

\begin{equation}
\begin{split}  
   \mathcal W_2=&\sum_{m=0}^{N}\sum_{n=0}^{N-m}|a_{mn}|^2\left(n-m\right)\\
   &+\sum_{m=0}^{N}\sum_{n=0}^{N-m}|a_{mn}||a_{m(n+2)}|\cos{(\beta_{mn}-\beta_{m(n+2)}-2\zeta(z))}\sqrt{(n+2)(n+1)}\\
  &-\sum_{m=0}^{N}\sum_{n=0}^{N-m}|a_{mn}||a_{(m+2)n}|\cos{(\beta_{mn}-\beta_{(m+2)n}-2\zeta(z))}\sqrt{(m+2)(m+1)}.
  \end{split}\label{w2}
\end{equation}

\begin{footnotesize}
\begin{equation}
\begin{split}  
   \mathcal W_2'=\frac{d\mathcal W_2}{dz}=&2\frac{z_r}{z^2+z_r^2}\sum_{m=0}^{N}\sum_{n=0}^{N-m}|a_{mn}||a_{m(n+2)}|\sin{(\beta_{mn}-\beta_{m(n+2)}-2\zeta(z))}\sqrt{(n+2)(n+1)}\\
  &-2\frac{z_r}{z^2+z_r^2}\sum_{m=0}^{N}\sum_{n=0}^{N-m}|a_{mn}||a_{(m+2)n}|\sin{(\beta_{mn}-\beta_{(m+2)n}-2\zeta(z))}\sqrt{(m+2)(m+1)}.
  \end{split}
\end{equation}
\end{footnotesize}

 \begin{equation}
\mathcal X_1=\sum_{m=0}^{N}\sum_{n=0}^{N-m}|a_{mn}||a_{(m+1)n}|\cos{(\beta_{mn}-\beta_{(m+1)n}-\zeta(z))}\sqrt{m+1}\label{x1}.
 \end{equation}
 
  \begin{equation}
\mathcal X_1'=\frac{d\mathcal X_1}{dz}=\frac{z_r}{z^2+z_r^2}\sum_{m=0}^{N}\sum_{n=0}^{N-m}|a_{mn}||a_{(m+1)n}|\sin{(\beta_{mn}-\beta_{(m+1)n}-\zeta(z))}\sqrt{m+1}.
 \end{equation}
 
 \begin{equation}
\mathcal X_2=\sum_{m=0}^{N}\sum_{n=0}^{N-m}|a_{mn}||a_{m(n+1)}|\cos{(\beta_{mn}-\beta_{m(n+1)}-\zeta(z))}\sqrt{n+1}\label{x2}.
\end{equation}

 \begin{equation}
\mathcal X_2'=\frac{d\mathcal X_2}{dz}=\frac{z_r}{z^2+z_r^2}\sum_{m=0}^{N}\sum_{n=0}^{N-m}|a_{mn}||a_{m(n+1)}|\sin{(\beta_{mn}-\beta_{m(n+1)}-\zeta(z))}\sqrt{n+1}.
\end{equation}

\begin{equation}
\mathcal  X_{mn}(\psi)=|a_{mn}||a_{(m+1)n}| \sqrt{m+1}\cos\psi.
\end{equation}

\begin{equation}
\mathcal  X_{\beta_{mn}}=\beta_{mn}-\beta_{(m+1)n}.
\end{equation}

\begin{equation}
\mathcal Y_{mn}(\psi)=|a_{mn}||a_{m(n+1)}| \sqrt{n+1}\sin\psi.
\end{equation}

\begin{equation}
\mathcal  Y_{\beta_{mn}}=\beta_{mn}-\beta_{m(n+1)}.
\end{equation}

\begin{equation}
\mathcal X_{s}(\psi)=\sum_{m=0}^N\sum_{n=0}^{N-m}\mathcal X_{mn}(\psi)\sin{\bigl(\mathcal  X_{\beta_{mn}}\bigr)}.
\end{equation}

\begin{equation}
\mathcal X_{c}(\psi)=\sum_{m=0}^N\sum_{n=0}^{N-m}\mathcal X_{mn}(\psi)\cos{\bigl(\mathcal  X_{\beta_{mn}}\bigr)}.
\end{equation}

\begin{equation}
\mathcal Y_{s}(\psi)=\sum_{m=0}^N\sum_{n=0}^{N-m}\mathcal Y_{mn}(\psi)\sin{\bigl(\mathcal  Y_{\beta_{mn}}\bigr)}.
\end{equation}

\begin{equation}
\mathcal Y_{c}(\psi)=\sum_{m=0}^N\sum_{n=0}^{N-m}\mathcal Y_{mn}(\psi)\cos{\bigl(\mathcal  Y_{\beta_{mn}}\bigr)}.
\end{equation}

\begin{equation}
\mathcal W_{mn_x}(\psi)=2|a_{mn}||a_{(m+2)n}| \sqrt{(m+1)(m+2)}\cos^2\psi.
\end{equation}

\begin{equation}
\mathcal W_{mn_y}(\psi)=2|a_{mn}||a_{m(n+2)}| \sqrt{(n+1)(n+2)}\sin^2\psi.
\end{equation}

\begin{equation}
\mathcal W_{mn_{xy}}(\psi)=4|a_{mn}||a_{(m+1)(n+1)}| \sqrt{(m+1)(n+1)}\sin\psi\cos\psi.
\end{equation}

\begin{equation}
\begin{split}
\mathcal W_{1}(\psi)=&\sum_{m=0}^N\sum_{n=0}^{N-m}\bigl(|a_{mn}|^2(2m\cos^2\psi+2n\sin^2\psi+1)\bigr.\\
&\bigl.+4|a_{m(n+1)}||a_{(m+1)n}|\sqrt{(m+1)(n+1)}\cos\psi\sin\psi\cos(\beta_{m(n+1)}-\beta_{(m+1)n})\bigr).
\end{split}
\end{equation}

\begin{equation}
\mathcal V_{mn_x}=\beta_{mn}-\beta_{(m+2)n}.
\end{equation}

\begin{equation}
\mathcal V_{mn_y}=\beta_{mn}-\beta_{m(n+2)}.
\end{equation}

\begin{equation}
\mathcal V_{mn_{xy}}=\beta_{mn}-\beta_{(m+1)(n+1)}.
\end{equation}

\begin{equation}
\mathcal W_{x_c}(\psi)=\sum_{m=0}^N\sum_{n=0}^{N-m}\mathcal W_{mn_x}(\psi)\cos\bigl(\mathcal V_{mn_x}\bigr).
\end{equation}

\begin{equation}
\mathcal W_{y_c}(\psi)=\sum_{m=0}^N\sum_{n=0}^{N-m}\mathcal W_{mn_y}(\psi)\cos\bigl(\mathcal V_{mn_y}\bigr).
\end{equation}

\begin{equation}
\mathcal W_{xy_c}(\psi)=\sum_{m=0}^N\sum_{n=0}^{N-m}\mathcal W_{mn_{xy}}(\psi)\cos\bigl(\mathcal V_{mn_{xy}}\bigr).
\end{equation}

\begin{equation}
\mathcal W_{x_s}(\psi)=\sum_{m=0}^N\sum_{n=0}^{N-m}\mathcal W_{mn_x}(\psi)\sin\bigl(\mathcal V_{mn_x}\bigr).
\end{equation}

\begin{equation}
\mathcal W_{y_s}(\psi)=\sum_{m=0}^N\sum_{n=0}^{N-m}\mathcal W_{mn_y}(\psi)\sin\bigl(\mathcal V_{mn_y}\bigr).
\end{equation}

\begin{equation}
\mathcal W_{xy_s}(\psi)=\sum_{m=0}^N\sum_{n=0}^{N-m}\mathcal W_{mn_{xy}}(\psi)\sin\bigl(\mathcal V_{mn_{xy}}\bigr).
\end{equation}

\begin{equation}
\mathcal Z_1(\psi)=P_{MEM}(\mathcal W_1(\psi)-\mathcal W_{x_c}(\psi)-\mathcal W_{y_c}(\psi)-\mathcal W_{xy_c}(\psi))-4(\mathcal X_s(\psi)+\mathcal Y_s(\psi))^2.
\end{equation}

\begin{equation}
\mathcal Z_2(\psi)=2P_{MEM}(\mathcal W_{x_s}(\psi)+\mathcal W_{y_s}(\psi)+\mathcal W_{xy_s}(\psi))-8(\mathcal X_s(\psi)+\mathcal Y_s(\psi))(\mathcal X_c(\psi)+\mathcal Y_c(\psi)).
\end{equation}

\begin{equation}
\mathcal Z_3(\psi)=P_{MEM}(\mathcal W_1(\psi)+\mathcal W_{x_c}(\psi)+\mathcal W_{y_c}(\psi)+\mathcal W_{xy_c}(\psi))-4(\mathcal X_c(\psi)+\mathcal Y_c(\psi))^2.
\end{equation}

\bibliographystyle{opticajnl}
\bibliography{sample}





\end{document}